\documentclass[11pt]{article}

\usepackage{jheppub}

\usepackage[utf8]{inputenc}
\usepackage{setspace}
\usepackage{verbatim}
\usepackage[usenames,dvipsnames]{xcolor}
\usepackage{amsmath}
\usepackage{latexsym}
\usepackage{tikz}
\usepackage{mathtools}

\renewcommand{\d}{\mathrm{d}}
\newcommand{\p}{\partial}

\newcommand{\Tr}{{\rm Tr }}
\newcommand{\tr}{{\rm tr }}

\makeatletter
\newcommand{\subalign}[1]{%
  \vcenter{%
    \Let@ \restore@math@cr \default@tag
    \baselineskip\fontdimen10 \scriptfont\tw@
    \advance\baselineskip\fontdimen12 \scriptfont\tw@
    \lineskip\thr@@\fontdimen8 \scriptfont\thr@@
    \lineskiplimit\lineskip
    \ialign{\hfil$\m@th\scriptstyle##$&$\m@th\scriptstyle{}##$\crcr
      #1\crcr
    }%
  }
}
\makeatother

\usepackage{amssymb}
\usepackage{graphicx}

    \newcommand{\beq}{\begin{equation}}
    \newcommand{\eeq}{\end{equation}}
    \newcommand\beqa{\begin{eqnarray}}
    \newcommand\eeqa{\end{eqnarray}}
\def\<{\left<}
\def\>{\right>}
\def\d{\partial}
\newcommand{\nn}{\nonumber}
\newcommand{\eq}[1]{(\ref{#1})}

\title{A Flow in the Forest}

\emailAdd{agorsky3$\bullet$gmail.com}
\emailAdd{kazakov$\bullet$lpt.ens.fr}
\emailAdd{fedor.levkovich$\bullet$gmail.com}
\emailAdd{mishnyakovvv$\bullet$gmail.com}

\author[a]{Alexander Gorsky,}
\author[b]{Vladimir Kazakov,}
\author[c,1]{Fedor Levkovich-Maslyuk,\note{Also at Institute for Information Transmission Problems, Moscow 127994, Russia}}
\author[d,e,f,g]{Victor Mishnyakov}

\affiliation[a]{
 Institute for Information Transmission Problems, Moscow 127994, Russia
 } 

\affiliation[b]{
 Laboratoire de Physique de l'\'{E}cole Normale Sup\'{e}rieure,
 CNRS, Universit\'{e} PSL, Sorbonne Universit\'{e}s,
 24 rue Lhomond, 75005 Paris, France
 }

 \affiliation[c]{
 Universit\'{e} Paris Saclay, CNRS,  CEA, Institut de physique th\'{e}orique,   91191, Gif-sur-Yvette, France
 } 

 \affiliation[d]{  Lebedev Physics Institute, Moscow 119991, Russia
 } 
 \affiliation[e]{  NRC ``Kurchatov Institute'', 123182, Moscow, Russia
 } 
 \affiliation[f]{ MIPT, Dolgoprudny 141701, Russia
 } 
 \affiliation[g]{Institute for Theoretical and Mathematical Physics, Lomonosov Moscow State University, Moscow 119991, Russia}

\abstract{Using the matrix-forest theorem  and the Parisi-Sourlas trick we  formulate and solve a one-matrix model with non-polynomial potential  which provides perturbation theory for massive spinless  fermions on dynamical  planar graphs. This is a lattice version of 2d quantum gravity coupled to massive spinless fermions.  Our model equivalently describes the ensemble of spanning forests on the same graphs. The solution is formulated in terms of an elliptic curve.  We then focus on a near-critical scaling limit when both the graphs and the trees in the forests are macroscopically large. In this limit we obtain  one-point scaling functions (condensates), parameterized in terms of  the Lambert function. Our results provide a rare example where one can explore the flow between two gravity models -- in this case, the theories of conformal matter coupled to 2d gravity with $c=-2$ (large trees regime) and $c=0$ (small trees regime). We also  compute the disc partition functions with Dirichlet and Neumann boundary conditions in the same critical limit.}

\begin{document}

\maketitle

\section{Introduction}

Matrix model approach to \(2d\)  quantum gravity, first proposed  in~\cite{Kazakov:1985ds,David:1984tx,Kazakov:1985ea},   has undergone since then     many important and diverse developments (see~   \cite{Kazakov:1988ch,DiFrancesco:1993cyw,Brezin:1994eb} for early 
results, and a more recent review~\cite{Anninos:2020ccj}). They range from  the study of critical exponents in the presence of conformal matter with a central charge \(c<1\)~\cite{Kazakov:1986hu,Boulatov:1986sb,Kazakov:1987qg,Kostov:1988fy,Kazakov:1989bc} to the construction of quantum   field theory for non-critical strings via the double scaling limit~\cite{Brezin:1990rb,Douglas:1989ve,Gross:1989vs,Douglas:1989dd}.  As was first proposed in~\cite{Kazakov:1988ch}, and then generalized to the double-scaling limit~\cite{Brezin:1989ss,Gross:1990ay,Ginsparg:1990as,Parisi:1989dka},  a  special, limiting \(c=1\) case of non-critical string theory can be studied via the matrix quantum mechanics (MQM)~\cite{Brezin:1977sv}. An  interesting peculiarity of \(c=1\)  case are the integer critical exponents,  resulting in the  logarithmic scaling, as well as in the presence of Berezinsky-Kosterlitz-Thouless  vortex excitations on the world-sheet of the \(c=1\) string~~\cite{Gross:1990md,Boulatov:1991fp,Boulatov:1991xz}, whose condensation can lead to the black hole formation~\cite{Kazakov:2000hea}. The \(c=1\) MQM model has served as a basis of formulation and study of type-0  \(2d\)   superstring models~\cite{Douglas:2003up}. The other interesting cases with integer critical exponents and a logarithmic scaling are various models with matter central charges $c=-2$ and $c\to 0$, in particular, the $Q\to 0,1$ limits of the $Q$-state Potts model~\cite{Kazakov:1987qg,Kazakov:1988fv,Daul:1994qy}\footnote{The solution of the saddle point problem for the matrix model of  Potts spins on random graphs was found by I.Kostov and V.K., published in~\cite{kostov1989random}} and the   $n\to 0,-2$ limits of $O(n)$ spin model\cite{Kostov:1987kt,Kostov:1988fy,Kostov:1991cg,Kostov:1992pn,Kostov:2006ry,Eynard:1992cn}\footnote{See also \cite{Eynard:1995nv,Eynard:1995zv}}, both on planar graphs. More recently the matrix 
models for  the pure JT gravity \cite{sss} and JT gravity coupled with the massive scalar \cite{jafferis}  have been
suggested.

Beyond the scaling properties,  certain important physical
quantities have been studied and computed for various central charges of the matter \(c\), such as the disc and cylinder partition functions and the multi-loop correlators for the minimal \((p,q)\) series, with \(c=1-6\frac{(p-q)^2}{pq}\)~\cite{Kostov:1991cg}. However, the questions of universal flows between  critical points with different central charges of the matter  hasn't yet been carefully studied.   A particular, well-studied case is the non-unitary \((2,2n-1)\) series~\cite{Belavin:2008kv,Belavin:2013nba}, where the precise map from the matrix model to the minimal Liouville theory of  \(2d \) gravity with conformal matter~\cite{Polyakov:1981rd,Knizhnik:1988ak,David:1988hj,Distler:1988jt},  completing and old proposal~\cite{Moore:1991ir}, was formulated. The study on the matrix side is simplified here by formulation in terms of the  one-matrix model   with a polynomial potential~\cite{Kazakov:1989bc,Staudacher:1989fy}. Certain flows have been observed between between various $p,q$  points within the $O(n)$ model of Kostov~\cite{Kostov:1991cg,Kostov:1992pn,Kostov:2006ry}. Apparently, these are only particular examples, and a thorough study of the whole variety of such flows, especially around logarithmic criticalities mentioned above, is largely missing.

The main subject of this paper is the study of such flows, and of the associated physical quantities (one point functions, disc partition functions) for an
interesting case of \(2d\) QG  coupled to  \(c_{M}=-2\)  matter,  with a logarithmic scaling.
 This system can be represented  as the 2d QG  in the presence of complex Grassmannian scalar matter field \(\theta(\sigma)\). Such model has been studied
in both the matrix model approach, where the sum over geometries is represented by the sum over planar graphs~\cite{Kazakov:1985ea,Boulatov:1986jd,Kostov:1987kt}, as well as in continuous Liouville field theory approach ~\cite{Klebanov:1990ip,Klebanov:1990sn,Edwards:1991jx}. 

Massive spinless fermions on the fixed non-fluctuating torus have been explored in \cite{david88}\footnote{See also \cite{zamolodchikov2006massive} for a related model of fermions coupled to gravity.}.
If the four-fermion interaction for massive fermions is added at a  particular value of four-fermion coupling this 
model on a fixed graph is identical to $q\rightarrow 0$ Potts model and $O(-2)$ sigma model \cite{Caracciolo:2004hz}.
The $q\rightarrow 0$ Potts model coupled to 2d gravity has been considered in \cite{Kazakov:1987qg,Caracciolo:2009ac,Bondesan:2016osa}
via matrix model representation. At  a  particular value of the four-fermion coupling
the model gets reduced to the statistical ensemble of unrooted spanning forests.
The matrix potential is non-analytic and a complicated pattern of phase
transitions has been discovered \cite{Bondesan:2016osa}.

In this work, we will study a  generalization of this \(c=-2\)  QG matrix model where the matter field  \(\psi(\sigma)\) is massive.
The disc partition function of this \(2d\) QG theory in continuum is given by the functional integral 
\begin{align}\label{continuousZ}
Z(\Lambda_0,\eta,m_0)=\int D g(\sigma)\int D \psi(\sigma)D \bar\psi(\sigma)\exp\Big[&-\int _{\mathcal{D}}d^2\sigma\sqrt{g}\left(g^{\alpha\beta}\p_\alpha\bar\psi \p_\beta\psi+m_0^2   \bar\psi\psi+\Lambda_0+\kappa R\right)\notag\\
   &-\eta_0 \int_{\mathcal{\mathcal{\p D}}}ds\,[g(\sigma(s))]^{1/4}\Big]
\end{align}
where \(g\) is the \(2d\) metric, \(\Lambda_0 \)  is the bulk cosmological constant,  \(\eta_0\)  is the boundary cosmological constant, \(R\) is the Gaussian curvature and \(\kappa=2-2g\) is the Euler characteristics which in our paper will correspond to the  topologies of the sphere \(\kappa=2\) or of the disc \(\kappa=1\).    At zero mass \(m=0\), the matter field has the conformal central charge \(c_{M}=-2\).

Using the matrix-forest theorem~\cite{kelmans} and Parisi-Sourlas trick~\cite{Parisi:1979ka} we will derive the
corresponding matrix model, with a specific non-polynomial potential, with two-parameters --  the bare mass of fermions and the bare cosmological
constant.
In our matrix formulation, the functional integral over metrics will be represented by the sum over \(\phi^3\) (i.e. trivalent) planar graphs, and the matter fields \(\psi_j\) will be placed at the sites \(j=1,2,\dots V\) of each graph of size \(V\), see eq.~\eqref{logzeta}.

 If one turns on the mass, at large distances (small cosmological constant) the matter is screened out and one "flows" in the infrared to the pure gravity \(c_M=0\) point, with a renormalized cosmological constant. The study of this flow by the matrix model tools  is the principal task of this paper. Solving the corresponding one-matrix model with a special, non-polynomial potential, we will establish the universal scaling functions describing the one-point function and the disc partition functions (with various boundary conditions). 

We should stress here that, whereas a lot is known  about the universal critical behavior of various physical quantities (correlation functions, disc and annulus partition functions, etc)  at a given central charge of conformal matter~\cite{Kostov:1988fy,Kostov:1991cg,Daul:1993bg,Daul:1994qy,David:1990ge,Eynard:1992cn,Eynard:1992sg,Kazakov:2004du}, the universal flows between the fixed points with various central charges are not that well studied. In this paper, we present our results for the one-point function and the disc partition functions  for such a flow, between the \(c=-2\)  and \(c=0 \)  fixed points. 

As an example, we present in the introduction the $t$-parametric representation of the one point function of the type \(\Phi_s=\langle\bar\psi\psi\rangle:\) (see section \ref{sec:1pt} for details\footnote{in the notation used there this correlator corresponds to ${\rm tr}\;\phi^3$}) 
\begin{align}\label{universalFi}
\Phi_s&\sim 2({\cal J}+{\rm const})(  t^2-2  t)+3   t^2-2  t^3
\end{align}
 where (in certain units)
\begin{align}\label{universalJ}
{\cal J}=  t-  \log (m_*^2  t)\,,\qquad ({\cal J}=\frac{\Lambda_* }{m_*^2 })\,
\end{align}
and $\Lambda_*,m_*^2$ are the lattice analogues of $\Lambda_0,m_0^2$ of \eqref{continuousZ}, and ${\rm const}$ in \eq{universalFi} indicates a numerical constant (given in section \ref{sec:1pt}).
This equation is closely related upon exponentiation to the Lambert function which has appeared previously
in  related contexts. It was found in \cite{okuyama} that the Lambert function corresponds
to the peculiar brane which provides  the generating function for the multiple boundaries
in the Airy limit of topological gravity at genus zero. Since the brane insertion shifts the closed 
moduli \cite{Aganagic_2005} it defines these shifts as well. Another example of the 
Lambert function providing the moduli shift has been identified in the topological string
on $CP_1$ coupled to topological gravity \cite{nekrasov}. In our study the Lambert function
will play a similar role.
Lambert function also emerges in various formulas related to asymptotics of Hurwitz numbers~\cite{Bouchard:2007hi} and as the spectral curve of the type B topological string on $C^3$
in the limit of  one-leg  infinite framing.

One can easily see that this scaling function interpolates between pure 2d QG \(c=0 \) regime~\(\Phi_s\sim (\Lambda_c-\Lambda)^{3/2} \)  when \(\Lambda\ll  m^2\) and  the 2d QG with \(c=-2\) matter regime ~\(\Phi_s\sim \Lambda^{2}\log\Lambda \)  when \(\Lambda\gg  m^2\), described by spanning trees on large planar graphs.        

In addition to \eq{universalFi}, we also computed another 1-pt function of a similar type, which turns out to differ only by the value of the additive shift of ${\cal J}$ in \eq{universalFi}. This suggests a universality-like property that would be interesting to explore further.

We will also derive two universal disc partition functions describing the flow between the \(c=-2\)  and \(c=0 \)  fixed points,  for massive spinless worldsheet fermions  with Neumann and Dirichlet boundary conditions. The results are given by equations \eqref{Gsing} and \eqref{Hsing}, respectively. In the limits mentioned above for the one-point function \(\Phi_s\), they reproduce the known asymptotic behavior at the \(c=-2\)  and \(c=0 \)  fixed points~\cite{Kostov:1992pn}.\footnote{In the cited paper, the author presents another example of a flow between the same critical points, which he derives from the $O(-2)$ model on planar graphs. Apparently, that flow is different, it leads to a different universal scaling function. }

If we drop off the planarity condition, the sum over the graphs is nothing but the random regular graph (RRG) ensemble
which enjoys some exact results in the limit of large number of nodes.
In particular we can utilize the famous Kesten-McKay(KM) distribution for the resolvent and spectral 
density of RRG ensemble ~\cite{kesten1959symmetric,mckay1981expected}. Recently the RRG has attracted a lot
of attention from the very different perspective which has nothing to do with the quantum gravity. It is considered as  the toy model for the Hilbert
space of the interacting many-body system and according to the conjecture ~\cite{levitov} the 
one-particle Anderson localization of spinless fermion  on RRG with diagonal disorder is equivalent to the many-body localization
in the physical space-time (see \cite{mirlin} for the recent review and references therein). It was argued that the fragmentation of the 
Hilbert space graph into  some number of "trees" is one of the key mechanisms of transition to the MBL phase
(see \cite{Moudgalya_2022} for review and references therein). 
It is not clear whether the problem we are solving in this work -- the massive spineless fermions on {\it planar} graphs -- is directly related to the abovementioned RGG ensembles of {\it generic} graphs. Possibly, the similar phenomena can occur in our model in  the appropriate double scaling limit, summing up the large graphs (near criticality) of all topologies~\cite{Brezin:1990rb,Douglas:1989ve,Gross:1989vs,Douglas:1989dd}.

The paper is organized as follows. In section~\ref{sec:def} we define our model of massive spinless fermions on the random
planar graphs. It can be equally considered as the theory of anticommuting bosons. In section~\ref{sec:derivation} we derive the
one-matrix model for our theory using two different strategies. First we generalize the Parisi-Sourlas trick for the
massive case and obtain the generalization of the matrix model for $c=-2$ theory. Secondly we shall utilize the
matrix-forest theorem \cite{kelmans} for the massive determinant of the graph Laplacian and develop a combinatorial derivation of the matrix model. This derivation is somewhat similar to the derivation of the matrix model in
\cite{Caracciolo:2009ac,Bondesan:2016osa} for a slightly different model of spinless fermions coupled to
2d gravity. In Section~\ref{sec:solution} we find the one-cut solution of this one-matrix model in the planar limit via the standard tools, in terms of the elliptic  curve. In section~\ref{sec:crl} we derive the critical curve which bounds the regime of validity for the one-cut solution in  the two-dimensional  parameter space of the model (related to the $(\Lambda,m^2)$ space of \eqref{continuousZ} in the critical regime).  In section~\ref{sec:1pt} we
consider in more detail the limit of small fermion mass and define a new double scaling regime in the space of two parameters. We will compute two 1-pt functions and find they are given by closely related scaling functions that interpolate between $c=-2$ and $c=0$ critical regimes and are related to the Lambert function. In section~\ref{sec:disc}  we derive 
the universal  disc partition functions of the model for the Dirichlet and Neumann boundary conditions (involving the parameter analogous to the boundary cosmological constant $\eta_0$ from \eqref{continuousZ}). In section~\ref{conclusions}  we summarize the main results of the paper and lastly section~\ref{sec:further} discusses  several possible topics  for future research. The appendices contain various technical details, while in appendix \ref{sec:lambert} we discuss the role of the Lambert function in our model and related contexts. We also attach a Mathematica notebook with some of the lengthy explicit results.

\section{Definition of the model}
\label{sec:def}

In the spirit of the discrete, random lattice  approach to \(2d\)  QG, we will construct a matrix model for which    the perturbative expansion for the free energy, combined with the \(1/N\) expansion, is given  as  the following sum over planar graphs $G$: 
\begin{align}\label{logzeta}
\log \zeta=\sum_{G}N^{2-2g}\lambda^{|G|}\int\prod_{i\in G}d^2\psi_i\prod_{<ij>\in G}e^{-(\bar \psi_i-\bar\psi_j)(\psi_i-\psi_j)}\prod_{i\in G}e^{-m^2\bar\psi_i\psi_i}
\end{align}
where in the first exponent the propagators \(e^{-(\bar \psi_i-\bar\psi_j)(\psi_i-\psi_j)}\)  mimic the kinetic term of fermions in the action~\eqref{continuousZ}, the last product introduces the mass of fermions, \(\log\lambda\) plays the role of bare bulk cosmological constant and  \(g\)  represents the genus of the discretized worldsheet. In this paper we will focus on the strictly planar case. We see that this   partition function looks as a lattice analogue of the partition function \eqref{continuousZ}.~\footnote{The boundary term for the disc partition function will be introduced later within the matrix model formalism}

Integrating in \eqref{logzeta} over  \(\psi\)'s we rewrite this  partition function in the form  
\begin{equation}\label{RGlogV}
Z\equiv \log \zeta=\sum_{G}\, N^{2-2g}\lambda^{|G|}\det[m^2+\Delta(G)]
\end{equation}  where \(\Delta=\mathbb{-Q}\,+A\) is the graph Laplacian  with \(\mathbb{Q}=\text{diag}\{q_1,q_2,\dots\}\)  where \(q_j\) are the valencies of vertices and $A$ is the adjacency matrix. In what follows we will use, for definiteness, the 3-valent graphs, i.e. \(q_i=3\).
For \(m^2=0\) the model was solved in \cite{Kazakov:1985ea,Boulatov:1986jd}  using the spanning trees representation and interpreted there as 2d QG with \(c=-2\) matter. We will generalize it to the \(m^2 \ne 0\) case and use for it the Parisi-Sourlas  approach  proposed in \cite{David:1985et,Kostov:1987kt}.

Notice that the partition function of the $c=-2$ model coupled to gravity is given by the sum over graphs of the determinant
of the graph Laplacian with the zero mode removed, which we denote by ${\det}'\Delta(G)$. Nicely, it is related in a simple way with the massive determinant~\footnote{The properties of this spectral determinant are discussed in Appendix~\ref{sec:spectral}} since
\beq
 {\det}'\Delta(G)=\dfrac{d}{dm^2} \det(\Delta(G)-m^2)_{m^2=0} \ .
\eeq
Hence we have an exact relation between the $c=-2$ and massive partition functions
\beq\label{c=-2det}
Z_{c=-2}(\lambda)= \dfrac{d}{dm^2}Z(\lambda,m^2=0) 
\eeq

The partition function \eqref{RGlogV} can be usefully rearranged via the 
generalization of the Kirchhoff matrix-tree theorem 
to the matrix-forest theorem for characteristic polynomial for the graph 
Laplacian which was obtained  in  \cite{kelmans} (see also \cite{david88})
\begin{equation}
   \det[m^2+\Delta(G)] = \sum_{F=(F_1\dots F_l)\in G}
  \prod_{i=1}^{l} m^2 V(F_i)
\end{equation}
where $V(F_i)$ is the number of nodes in the tree $F_i$ and $m^2$ is the generating parameter for
the number of trees in the forest. That is, our model can be considered as
the partition function of  spanning rooted forests interacting with 2d gravity. 

Notice that the work \cite{Caracciolo:2009ac} considered a related model which involves massive spinless fermions
supplemented with the four-fermion term coupled to 2d gravity. In that case 
the generalization of the matrix-forest theorem exists \cite{Caracciolo:2004hz} but the model,
contrary to our case, reduces to the spanning \textit{unrooted} forests on fluctuating surface. Instead we have \textit{rooted} forests and no four-fermion interaction.

Since the degrees of all nodes are equal, our model can be considered as a particular case of the random regular graph (RRG) ensemble with general exponential weight:
\begin{equation}\label{RGmodel}
Z=\sum_{RRG}\, \exp\left(-\Tr V[A(G)]\right)
\end{equation}
where the sum goes w.r.t. graphs G with fixed number of nodes  with the same degree. This partition function describes the microcanonical ensemble with fixed area of the surface,  contrary to the canonical ensemble with the cosmological constant. Usually in consideration of RRG ensemble no planarity condition for the graphs  is implied. Since \(\Tr\) here is taken w.r.t. indices of adjacency matrix of a graph,  \(\Tr A^n\) gives the number of cycles of length \(n\) on the graph.
In our case,  the potential  will have a specific, determinant form \(V(A)=-\log((m^2-3)\,\mathbb{I}-A)\), and furthermore we restrict to only planar graphs among the whole 
3-valent RRG ensemble.

\section{Derivation of the matrix model}
\label{sec:derivation}

In this section we will derive the Hermitian one-matrix model which gives our partition  \eqref{logzeta} and provides
the interpolation between pure gravity and $c=-2$ conformal matter coupled
with 2d quantum gravity. We will use two different strategies which 
generalize two approaches used for the $c=-2$ case. First, extending the approach 
of \cite{Kostov:1987kt,david88} we will apply a massive version of the Parisi-Sourlas trick. 
Second, extending the  derivation of the matrix model for the unrooted spanning
forests in \cite{Bondesan:2016osa}\footnote{See also \cite{abdesselam2004grassmann}} we will derive the matrix model for our case of
spanning forests. We will show that the matrix model potentials we find in the two derivations coincide.

\subsection{Derivation via Parisi-Sourlas representation}\label{sec:PS}

Using the Parisi-Sourlas approach \cite{Parisi:1979ka,Kostov:1987kt,Klebanov:1990sn,Kazakov:1990nd}  we can represent our partition function as a super-matrix integral of the form
 \begin{equation}\label{partitionF}
\zeta=\int D^{N^2}\Phi(\psi)\,\,e^{NS(\Phi)}=\int d^{N^2}\phi \,d^{2N^2} \,d^{2N^2}\epsilon\,\,e^{NS(\Phi)}
\end{equation}
 where the functional integration is performed w.r.t. the matrix superfield 
 \begin{equation}\label{superfield}
\Phi(\psi)=\phi+\bar\psi\theta+\psi\bar \theta+\bar\psi\,\psi\epsilon\, \ ,
\end{equation}
where $\bar\theta,\theta$ are the grassmann valued matrix fields,  $\bar\psi,\psi$ are anticommuting parameters and $\epsilon$ and $\phi$ are commuting matrix fields. For our case, the matrix model potential producing the Feynman expansion \eqref{logzeta} with cubic $(q=3)$ vertices is given by:
\begin{align*}
S(\Phi)&=\frac{1}{2}\tr\int d^2\psi\int d^2\chi\,\Phi(\psi)\exp\{-(\bar \psi-\bar\chi)(\psi-\chi)\}\Phi(\chi)+\\ &+\frac{\lambda}{3}\int d^2\psi \,e^{-m^2\bar\psi\psi}\tr \left[\Phi(\psi)\right]^3
\end{align*}
where \(d^2\psi= d\psi\bar{\psi}\). The superfield propagator is chosen to reproduce the kinetic term of the worldsheet fermion action, while the coefficient of the cubic vertex is responsible for the mass term. One can rewrite the action in a more standard form by noticing that the propagator is a function of the grassmanian distance
\begin{align*}
\langle\Phi(\psi)\Phi(\chi)\rangle_{0}=\exp\{-(\bar \psi-\bar\chi)(\psi-\chi)\} =1-(\bar \psi-\bar\chi)(\psi-\chi)
\end{align*}
which can be inverted as
\begin{equation}
    (-1-\partial_\psi \partial_{\bar\psi}) \exp\{-(\bar \psi-\bar\chi)(\psi-\chi)\} = (\bar \psi-\bar\chi)(\psi-\chi) =\delta^{(2)}\left(\psi-\chi \right) \ .
\end{equation}
This produces the following action:
\begin{align}\label{DerAction}
S(\Phi)&=\tr\int d^2\psi\,\left(-\frac{1}{2}\Phi^2(\psi)-\frac{1}{2}\p_{\psi}\Phi(\psi)
\p_{\bar\psi}\Phi(\psi)+\frac{\lambda}{3}e^{-m^2
\bar\psi\psi}\left[\Phi(\psi)\right]^3\right) \ .
\end{align}
Now, we would like to integrate our the Grassmanian variables to obtain a bosonic matrix model, that would generate the same expansion \eqref{logzeta}. Taking the integrals
we obtain:
\begin{equation}\label{action}
S(\Phi)=\tr\,\left[\frac{1}{2}\epsilon^2-(\phi-\lambda\phi^2)\epsilon-\frac{1}{3}\lambda m^2\phi^3+\bar\psi\psi-2\lambda \bar\theta\theta\phi \right].
\end{equation}
After integration over auxiliary fields  \(\epsilon\) and \(\theta\) we get
\begin{equation}
\zeta=\int d^{N^2}\phi\,\det(1-2\lambda\phi)  \,\,\,e^{N\tr\,\left[-\frac{1}{2} (\phi-\lambda\phi^2)^2+\frac{1}{3}\lambda m^2\phi^3]\right]}
\end{equation}
and finally doing a change of variables   \(X=\phi- \lambda \phi^2\) we come to the one-matrix model partition function:
\begin{equation}\label{Xaction}
    \zeta = \int d^{N^2}X  \exp\left( N \tr \left[ -\frac{1}{2} X^2 + \dfrac{\lambda m^2}{3} (\phi(X))^3 \right] \right)
\end{equation}
with $\phi$ related to $X$ by
\beq\label{defphi}
\phi(X)=\frac{1}{2\lambda}\left(1-\sqrt{1-4\lambda X} \right) \ 
\eeq 
 where we have selected the proper root of the quadratic equation.
Thus we see that the result is a 1-matrix model with a particular non-polynomial potential. This opens the way to solve the model using standard techniques. In the next subsection we will provide yet another derivation leading to this matrix model.

Instead of $(\phi(X))^3$ interaction in the potential in~\eqref{Xaction} we could take any polynomial potential. Then, instead of trivalent graphs, we will study the same problem of spinless massive fermions (or forests) on the corresponding collection of planar graphs.   For the generic couplings of this potential the critical behavior we study below should not change (due to universality). However, we can have multicritical points of the kind known from~\cite{Kazakov:1989bc,Kostov:1992pn}.

\subsection{Combinatorial derivation of the matrix model}

Here we will present another, combinatoric, derivation of the matrix model \eq{Xaction} describing our partition function.

Let us consider the graphical interpretation for our action \eqref{Xaction} and demonstrate that it indeed reproduces the expansion \eqref{logzeta}. First, recall that the function \eq{defphi} appearing in our matrix model potential 
\begin{equation}
    \phi(X)=\frac{(1-\sqrt{1-4\lambda X})}{2\lambda}=X+\lambda  X^2+2 \lambda ^2 X^3+5 \lambda ^3 X^4 + \ldots = \sum_{n=1}^{\infty}\frac{1}{n+1}\binom{2 n}{n} \lambda^{n} X^{n+1}
\end{equation}
is the generating  function of rooted trees  with node degree \(q=3\). Here the expansion coefficients \(C_n=\frac{1}{n+1}\binom{2 n}{n}  \) are nothing but the Catalan numbers. Each vertex is weighted with the coupling \(\lambda\), and the exterior branches are decorated with the matrix \(X\) (`Christmas Tree'). In our potential we have \(\phi^3(X)\) which generates three such trees growing from the same vertex (call it `root vertex').

Next, we expand the exponent in \eqref{Xaction} under the matrix integral  w.r.t. the 2nd term in the action. This expansion generates forests of such Christmas trees with marked vertex. Each marked vertex (corresponding to \(\phi^3(X) \) is weighted with \(\lambda m^2\), while the other vertices are weighted with $\lambda$. Hence each such forest has weight  \(\lambda^{|G|} m^{2l}\), where $|G|$ and $l$ denote the total number of vertices and the number of trees respectively.

As a next step, we see that the Gaussian integral over \(X\) connects, via rainbow diagrams,  these  Christmas trees  into  graphs, so that we sum up over all ``forests"\ of such trees on each graph.
 All  graphs are planar, due to the matrix structure. 
We notice that this statistical-mechanical system, on each graph,  is the same as given  by the equation~(13) of \cite{Caracciolo:2004hz}, where it is proven to be equal to the  determinant \(\det[ m^2+\Delta(G)]\) on each graph \(G\). This is another confirmation of our derivation of the matrix model \eqref{Xaction}    in section \ref{sec:PS} and its relation to the partition function \eqref{RGlogV}. 
\\

To prove that the combinatorial factors, such as the Catalan numbers and symmetry factors of Feynman graphs in \eqref{Xaction}, do indeed match up with the Kirchoff theorem (the proof of its ``massless" version is given in Appendix~\ref{sec:matrix-tree}) and the expansion \eqref{RGlogV} we use the result of the paper \cite{Caracciolo:2009ac} which
 considers a model that generates forests of ''unrooted trees", i.e trees without a marked point. Instead of
\eqref{Xaction} it corresponds to a similar potential: 
\begin{equation}\label{ITALIANaction}
S_{X}=\tr\,\left[-\frac{1}{2} X^2+m^2X^2 \tilde V(\lambda X)\right]\,
\end{equation}
where 
\begin{align}
\tilde V(z)&=\frac{1}{12z^2} \left(-6 z^2+(1-4 z)^{3/2}+6 z-1\right) \notag\\
&=\,\sum _{n=1}^{\infty } \tilde c_n z^{n+2}\,,\qquad \text{where}\ \,\,\tilde c_n=\frac{(2 n)!}{n! (n+2)!} \ .
\end{align} 
 The coefficients of the expansion are equal to the Catalan numbers  \(C_n=\frac{(2 n)!}{n! (n+1)!}\) divided by \(n+2\) giving the number of unrooted trees with \(n\)  vertices with cyclic symmetry factored out. Notice now the following identity relating our potential with  \(\tilde V(z)\):
 \begin{equation}
\frac{\lambda}{3}\left(\frac{1-\sqrt{1-4\lambda X}}{2\lambda}\right)^3=\lambda\partial_\lambda \left(X^2 \tilde V(\lambda X )\right) \ .
\end{equation}
In terms of coefficients of expansion for the l.h.s.  this relation looks as follows: 
\begin{equation}
c_{n  }=n\frac{C_n}{(n+2)} \ .
\end{equation}
In other words, the number of  trees with \(n\)  marked vertices  is equal to the number of rooted trees divided by the cyclic permutation order \(n+2\) of external legs (called ``leafs" in \cite{Caracciolo:2009ac}), which makes these trees unrooted, and multiplied by the \(n\) -- the number of ways to mark one vertex.  This is another proof that our model \eqref{Xaction} describes indeed the sum over forests of rooted trees over planar graphs, which is equal, according to the above-mentioned theorem from  \cite{Caracciolo:2004hz}~(see equation (13) there), to the partition function of massive spinless fermions (our \(\psi_i\)'s in \eqref{logzeta}) on planar graphs.  In our notation, for each individual planar graph this theorem states 
\begin{equation}
\det[m^2+\Delta(G)]=\sum_{k=1}^{\infty} m^{2k} F_k(G) 
\end{equation}
where \(F_k(G)\)  is the number of forests consisting of \(k\) unrooted trees, each with one marked vertex, on this graph \(G\). 

\subsection{Boundary conditions for disc partition functions}
\label{sec:treeexp}
\label{sec:bound}

As announced in the Introduction, we are going to be interested in computing disc partition functions. There are two types of disc partition functions we can compute in the matrix model \eqref{Xaction}. One corresponds to a boundary with the $X$-matrix, and another to the matrix $\phi(X)$ that generates trees.
They correspond to two resolvents we will study: the resolvent $G(x)$ which generates $\langle\tr\; X^k\rangle$ correlators,
\beq
     G(x)=\sum_{k=0}^\infty \frac{1}{x^{k+1}}\frac{1}{N}\langle\tr\; X^k\rangle
\eeq
and the resolvent for $\langle\tr\;\phi^k\rangle$ correlators,
\beq
    H(r)=\sum_{k=0}^\infty \frac{1}{r^{k+1}}\frac{1}{N}\langle\tr\; \phi^k\rangle
\eeq
with $k$ in the correlator being the disc boundary length, while $\log x$ and $\log r$ play the role of the bare boundary cosmological constant.

One can expect that these two resolvents correspond to two different types of boundary conditions. We would like to formulate these boundary conditions in terms of the worldsheet action \eqref{continuousZ} for the anti-commuting bosons. To do this we first recall how the trees appear from the miscroscopic model on graphs \eqref{logzeta}.

Let us fix a graph $G$ and consider the action for this graph. To produce trees we expand the kinetic terms from \eqref{logzeta}. Keeping in mind that our variables are anti-commuting we obtain:
\begin{equation}
   \prod_{<i,j>} e^{|\psi_{ij}|^2} = \prod_{<i,j>} \left(1+|\psi_{ij}|^2 \right) \ .  
\end{equation}
Expanding the product over edges for each edge we can either choose to include it into the tree, which corresponds to choosing the  $|\psi_{ij}|^2$ term, or not, in which case we take the identity. Loops are forbidden because of the vanishing of the  cyclic product $\prod\limits_{<i,j> \in L}|\psi_{ij}|^2$. Hence we obtain the expression for trees $T$ on the graph:
\begin{equation}
     \prod_{<i,j>} e^{|\psi_{ij}|^2} = \sum_{T \in G} V(T) \prod_{<i,j> \in T} |\psi_{ij}|^2
\end{equation}
where $V(T)$ is the number of vertices in a tree. However in order to obtain a non-zero expression we should also introduce boundaries for the trees. The first type of boundary corresponds to insertions of $\Tr\; \phi^n$. In graph terms this means that we have a graph with trees that start from the boundary and some trees in the bulk. Trees in the bulk are generated by the mass terms, while boundary trees should be enforced by fixing the field $\phi$ on the boundary, as the tree-generating formula~\eqref{defphi} suggests,. This is done by inserting a specific boundary term under the integral:
\begin{equation}
   \left(\prod_{<i,j>} e^{|\psi_{ij}|^2}  \prod_{<b,j>} e^{|\psi_{bj}|^2} \right) \psi_b \bar{\psi}_b
\end{equation}
where $b$ denotes the boundary. In other words we fix the value of $\psi$ on the boundary.
\\\\
The other type of boundary corresponds to operators $\Tr\; X^n$. Now we have a boundary with edges going into the bulk, while trees appear already in the bulk and do not touch the boundary. One can describe such behaviour on the worldsheet by an insertion
\begin{equation}
   \prod_{<i,j>} e^{|\psi_{ij}|^2}  \prod_{<b,j>} e^{|\psi_{bj}|^2} \prod |\psi_{bj}|^2  
\end{equation}
effectively fixing the derivative of the field on the boundary.

Having described the discrete picture it is natural to conjecture that the proper continuous boundary conditions are Dirichlet conditions
\begin{equation}
    \psi(\partial D) = 0
\end{equation}
for the $\phi$-type boundary disc partition function $H(r)$  and Neumann conditions
\begin{equation}
    \left.\partial_{\perp} \psi \right|_{\partial D}=0
\end{equation}for the $X$-type boundary appearing in $G(x)$. 

The critical behaviour of both disc partition functions, related to the continuous limit of $2d$ QG interacting with massive fermions,
appears to be different. We will compute it in section \ref{sec:disc}.

\section{Solution of the matrix model in the planar limit: saddle point}
\label{sec:solution}

As discussed above, the problem we study reduces to a 1-matrix model with the potential
 \beq
 \label{pote}
     V(x)=\frac{1}{2}x^2-\frac{3}{16} \frac{M}{\lambda^2} \left(1-\sqrt{1-4\lambda x}\right)^3
 \eeq
where we make a redefinition of the mass parameter for convenience:
\beq
    M=\frac{2}{9}m^2
\eeq
In this section we describe how to solve this matrix model at large $N$ and present the 1-cut solution in detail. 

\begin{figure}
    \centering
    \includegraphics[scale=0.6]{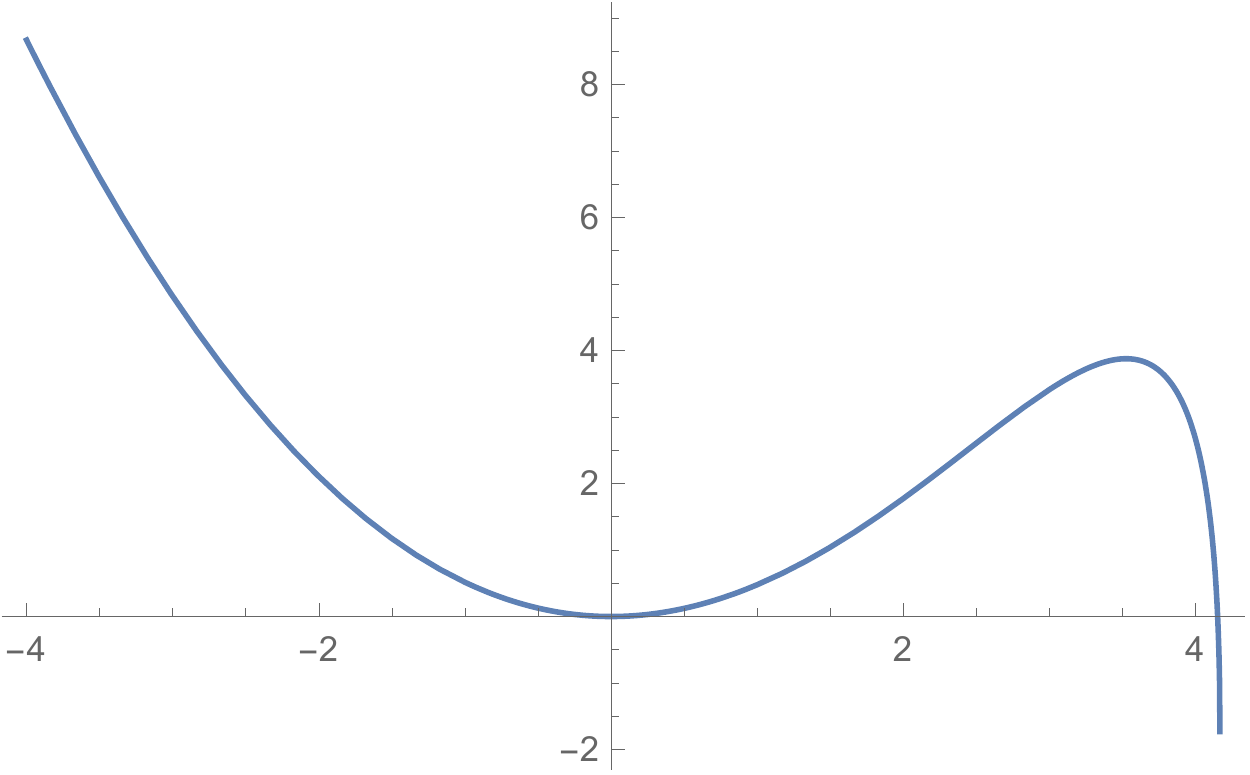}
    \caption{The matrix model potential \eq{pote} at  $\lambda=6/100\ , \ M=1/15$.}
    \label{fig:pot}
\end{figure}

We give a plot of the potential  on figure \ref{fig:pot}.  One can check that the shape remains the same regardless of the values of $M$ and $\lambda$ (with $M,\lambda>0$) -- namely, as we go from negative $x$ the potential has a local minimum at $x=0$ and then a local maximum before falling off to some finite value at $x=1/(4\lambda)$ which is the boundary of the allowed region due to the presence of the  square root in the potential.

Since the potential has a local minimum at $x=0$, it can be populated by eigenvalues and thus it is natural to look for a 1-cut solution which can be found by standard methods and which we will focus on in this paper. We parameterize the endpoints of the cut by $b$ and $a$ with $b<a$ and also denote the branch point of the square root by\footnote{We hope this notation will not create confusion with the notation $c$ for the central charge of the model}
\beq
    c=\frac{1}{4\lambda}
\eeq
so that $b<a<c$ are the three branch points in our problem (with one more at infinity coming from the square root) meaning it is resolvable in elliptic functions. Introducing the density of eigenvalues $\rho(x)$ normalised to $1$ on the interval $[b,a]$ and the resolvent
\beq
    G(x)=\int_b^a\frac{dy\;\rho(y)}{x-y}
\eeq
we can write the saddle point equation on the $[b,a]$ cut as
\beq
    G(x+i0)+G(x-i0)=x+9cM+\frac{9}{2}M\sqrt{c}\frac{x-2c}{\sqrt{c- x}} \ .
\eeq
The solution can be written easily as
\begin{align}
\label{Gasint}
G(x)=\sqrt{x-a}\sqrt{x-b}\int_b^a \frac{dy}{2\pi( x-y)}  \frac{1}{\sqrt{(a-y)(y-b)}}\left(y+9cM+\frac{9}{2}M\sqrt{c\,}\,\,\frac{y-2c}{\sqrt{c- y}}\right) \ .
\end{align}
The integrals here can be taken explicitly (see appendix \ref{app:ell}) and we find the resolvent,
\beqa &&
\label{Gres1}
    G(x)=\frac{9}{4\pi} \sqrt{c} M \sqrt{x-a}\sqrt{x-b}
   \left(\frac{2 (x-2 c) \Pi
   \left(\left.\frac{a-b}{x-b}\right|1-p\right)}{\sqrt
   {c-b} (x-b)}-\frac{2 K}{\sqrt{c-b}}\right)
    \\ \nn &&
   +\frac{1}{2}
    \left(-\sqrt{x-a}\sqrt{x-b}+9c M+x\right)
\eeqa
where we denoted
\beq
    p=\frac{c-a}{c-b}
\eeq
and $K\equiv K'(p)\equiv K(1-p)$ is the elliptic integral with modulus\footnote{We denote the modulus by $1-p$ for convenience, so that $p\to 0$ corresponds to the modulus approaching $1$ which is the limit we will study later. In Mathematica the functions we use correspond to EllipticK[1-p], EllipticPi[n,1-p], etc.} $1-p$. 

Then we can find the density as the discontinuity of the resolvent on the $[b,a]$ cut,
\beq
\label{Grho}
    G(x\pm i\epsilon)-G(x-i\epsilon)=-2 i\pi \rho(x)
\eeq
which gives the density in terms of elliptic functions as well\footnote{in terms of the expression in \eq{densityreg}, the density we should integrate to get $G$ is to be taken as $\rho(x+i\epsilon)$}, \beqa \label{densityreg}
    \rho(x)&=&\sqrt{(a-x)(x-b)}\left(\frac{9 \sqrt{c} M  K}{2 \pi ^2 \sqrt{c-b}}+\frac{1}{2\pi}\right)
     \\ \nn 
     &-&\frac{9 \sqrt{c} M
   (x-2 c) }{2 \pi ^2 \sqrt{c-b}}\sqrt{\frac{a-x}{x-b}}\left(K-\Pi\left(\frac{x-b}{c-b},1-p\right)\right) \ .
\eeqa

\subsection{Fixing parameters}

We  fix \(a,b\)  from the condition that for $x\to\infty$ we must have \(G(x)\simeq0\cdot x^0+\frac{1}{x}+\mathcal{O}(x^{-2}) \) which follows from the definition of the resolvent and the normalisation of the density. 
Introducing the variable
\beq
    y=\frac{\sqrt{c-b}}{\sqrt{c}} 
\eeq
we get two relations fixing $y$ and $p$ (or equivalently the branch points $a$ and $b$) in terms of our original couplings $\lambda$ and $M$, namely  the $x^0$ term in $G(x)$ gives
\beq
\label{x0res}
     -\pi(p+1)\,y^3-18 M E\, y^2+2 \pi  \left(9 M+1\right)\,y-18  M K=0
\eeq
relating $p$ and $y$, while from the $1/x$ term we have
\begin{equation}
 \label{x1res}
-\lambda^2 + \frac{(1-p)^2}{256}\,y^4  -\frac{3 M}{64\pi} (-(p+1) E+2p K)\,y^3+\frac{9 M}{64\pi} (-(p+1) K+2 E)\, y=0 \ .
\end{equation}
Here we denoted the elliptic integral of the 2nd kind as  $E\equiv E'(p)\equiv E(1-p)$. Notice that $\lambda$ only enters the second equation and is explicitly expressed as a function of $y,p$.

As these are two polynomial equations in $y$, we can exclude $y$ by taking their resultant\footnote{We recall that for two polynomials with roots $a_i$ and $b_j$ the resultant is defined as $\prod_{i,j}(a_i-b_j)$.  As a symmetric function in both sets of roots, it is a polynomial in the coefficients of the original two polynomials. It is also implemented in Mathematica.} which we denote as $P(\lambda,M,p)$, so we have
\beq
\label{cmM}
    P(\lambda,M,p)=0 \ .
\eeq
This gives a lengthy, though explicit, equation linking $\lambda,M$ and $p$. The function $P$ is a polynomial of 6th order in $M$ and of 3rd order in $\lambda^2$, thus one can in principle write $\lambda(M,p)$ explicitly.

\begin{figure}
    \centering
    \includegraphics[scale=0.65]{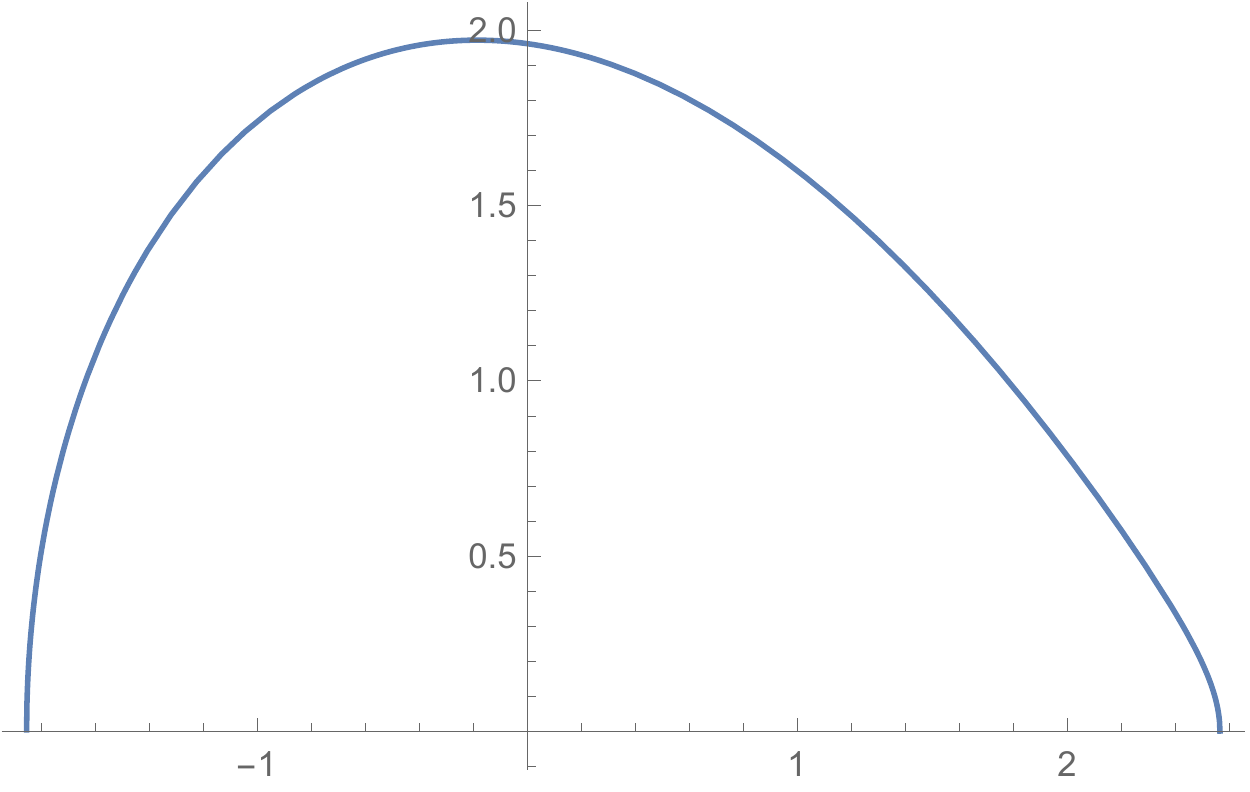}
    \caption{The density $\rho(x)$ given by \eq{densityreg} on the $[b,a]$ cut for $\lambda=5/106\approx 0.047\ , \ M=1156/25\approx 46$. We chose the small value of $\lambda$ and large $M$ so that the density has a nontrivial profile substantially different from a semicircle.}
    \label{fig:rho}
\end{figure}

These equations are a complete system that allows one to obtain the 1-cut solution at a given value of our parameters $\lambda$ and $M$. In order to e.g. solve the system numerically, we can first solve \eq{cmM} for $p$ and then plug the result into \eq{x1res} which is then solved for $y$. As an example we give a plot of the density in figure \ref{fig:rho}. Note that in general these equations have multiple solutions and we should be careful to select the `physical' one, for which the density is real and positive. We discuss this requirement in more detail in section \ref{sec:crl}. 

\begin{figure}
    \centering
    \includegraphics[scale=0.6]{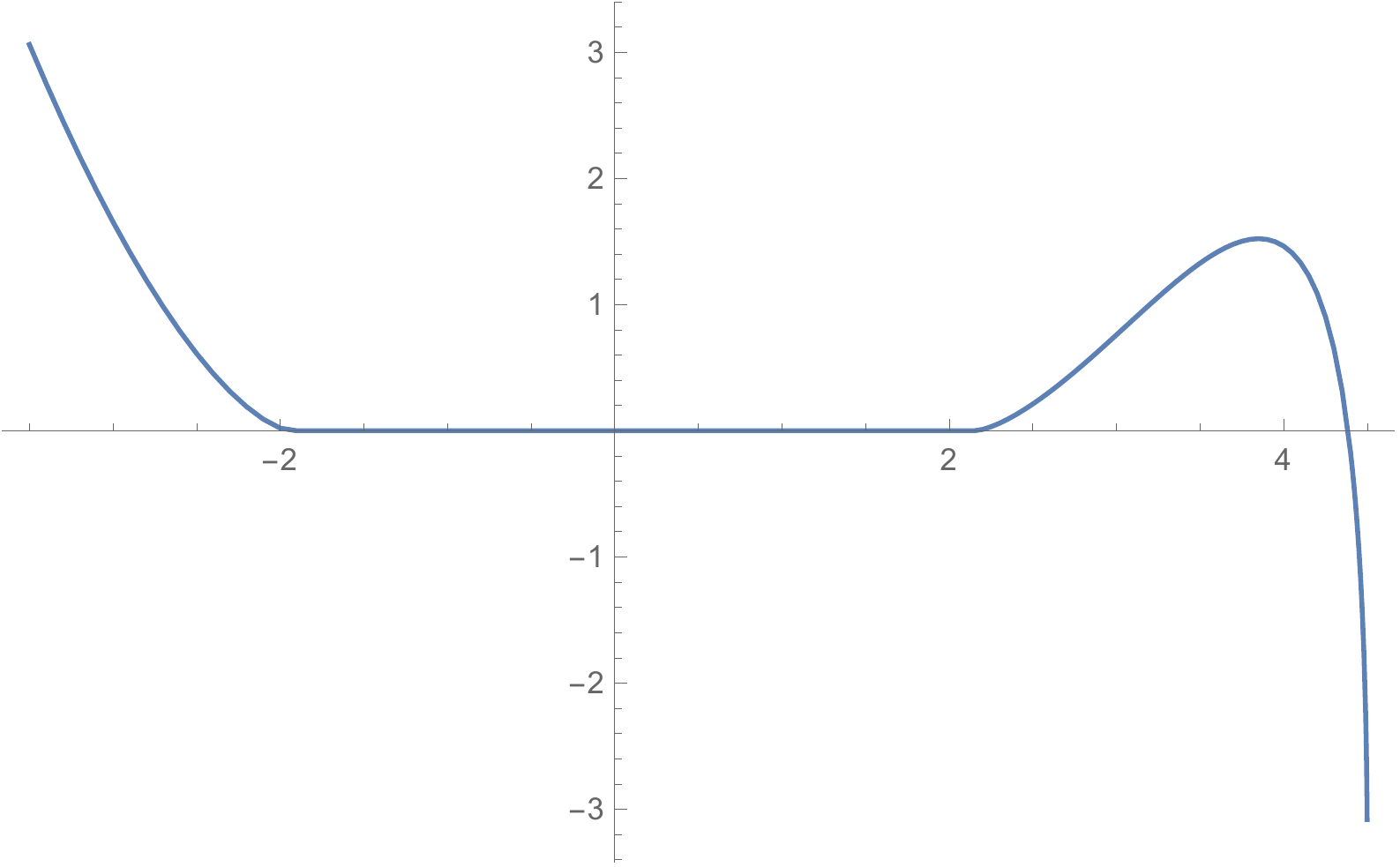}
    \caption{The effective potential $V_{eff}(x)$ at  $\lambda=1/18\ , \ M=16/100$.}
    \label{fig:Veff}
\end{figure}

It is instructive to plot the effective potential $V_{eff}$ felt by the eigenvalues, which as usual is given by
\beq
    V_{eff}(x)=V(x)-2\int_0^xdy(G(y-i0)+G(y+i0))
\eeq
From this definition it follows that  it has a flat section on the $[b,a]$ cut where it vanishes. We show a plot of it on figure \ref{fig:Veff}. Apart from the flat region, its shape is similar to the original matrix model potential.

\subsection{Exact results for correlators}
\label{sec:excorr}

Let us discuss how to compute the observables in this model. First, the correlators $\langle \frac{1}{N}\tr \; X^k \rangle$ are encoded in the large $x$ expansion of the resolvent $G(x)$ and can be obtained directly by expanding \eq{Gres1}. Another important set of observables are built from the matrix $\phi$ related to $x$ by \eq{defphi}. They can be written as
\begin{equation}
\label{phii1}
    \langle \frac{1}{N}\tr\;\phi^k \rangle = \int_b^a dx\;\rho(x) \left(2c \left(1-\sqrt{1-\dfrac{x}{c}} \right) \right)^k \ , \ \ k=1,2,\dots
\end{equation}
These integrals are highly nontrivial since the density involves the elliptic $\Pi$ function. However, remarkably, we found a way to compute them in closed form for any given $k$.
The tricky part is the integration of the square root with the elliptic integral of third kind in \eqref{densityreg}.  To explain how this integral can be evaluated explicitly, first denote
\begin{equation}
 n=\dfrac{b-x}{b-c} \ .
\end{equation}
The integral is then over $n$ between $0$ and $1$. The relevant piece is:
\begin{equation}
    \int_{0}^1 dn (n-1)^{k/2} \sqrt{\frac{1-p-n}{n}} (b n-b-c n+2 c) \Pi (n|1-p) 
\end{equation}
The key trick is to get rid of the integration of elliptic functions, by using the identities
presented in Appendix A (see section \eq{app:eint} for details).
The integration of $\Pi(n|1-p)$ is reduced to an integral of a derivative, while the remaining part contains only algebraic dependence on $n$ and can be easily integrated.

As an example, important for the further computations of this one point function in the critical regime, we give below the explicit result for $k=1$
\beqa
\nn
    \langle \frac{1}{N}\tr\;\phi \rangle&=&\frac{c^3 y}{2 \pi ^2}\left[  \frac{4}{15} \pi  p (p+1) y^4 K-
    -
    \frac{8}{15} \pi  \left(p^2-p+1\right) y^4 E+\frac{1}{4} \pi ^2 (p-1)^2 y^3 \right.
   \\ \nn &&
  +M \left( 
 K \left(12 y \left(p y^2+p+1\right) E-3 \pi  \left(p \left(2 y^2+3\right)+3\right)\right)
  \right.-
   \\ \nn &&
   \left. \left.
   - 6 p y K^2-6 y \left((p+1) y^2+3\right)
   E^2+3 \pi  \left((p+1) y^2+6\right) E
   \right) \right] \ .
\eeqa
We also give the result for $\langle \frac{1}{N}\tr\;\phi^3 \rangle$, which is more lengthy, in a Mathematica notebook accompanying this paper. This observable is particularly important as it gives the derivative of the partition function w.r.t. the mass $m^2$, which in view of \eq{logzeta} means it describes the fermionic condensate of the type $\langle \bar\psi \psi\rangle$.

It would be also interesting to obtain an explicit result similar to \eq{Gres1} for the generating function of these correlators, which seems to be quite challenging but might be possible to do by using the methods of \cite{Kostov:2006ry}.

\section{The critical line}
\label{sec:crl}

The critical regime in our model, corresponding to large graphs, is obtained by going close to a singularity of the partition function. A (standard) shortcut to deriving the condition for criticality is requiring that the density has zero slope at the endpoint, $\rho'(a)=0$. The reason for this is that generically we have $\rho(x)\sim \sqrt{a-x}$ when $x\to a$, and since the density should be positive the coefficient in front of the square root must be nonnegative as well. The case when the coefficient becomes zero thus belongs to the boundary of the allowed parameter space where the model becomes singular.

\begin{figure}
    \centering
    \includegraphics[scale=0.6]{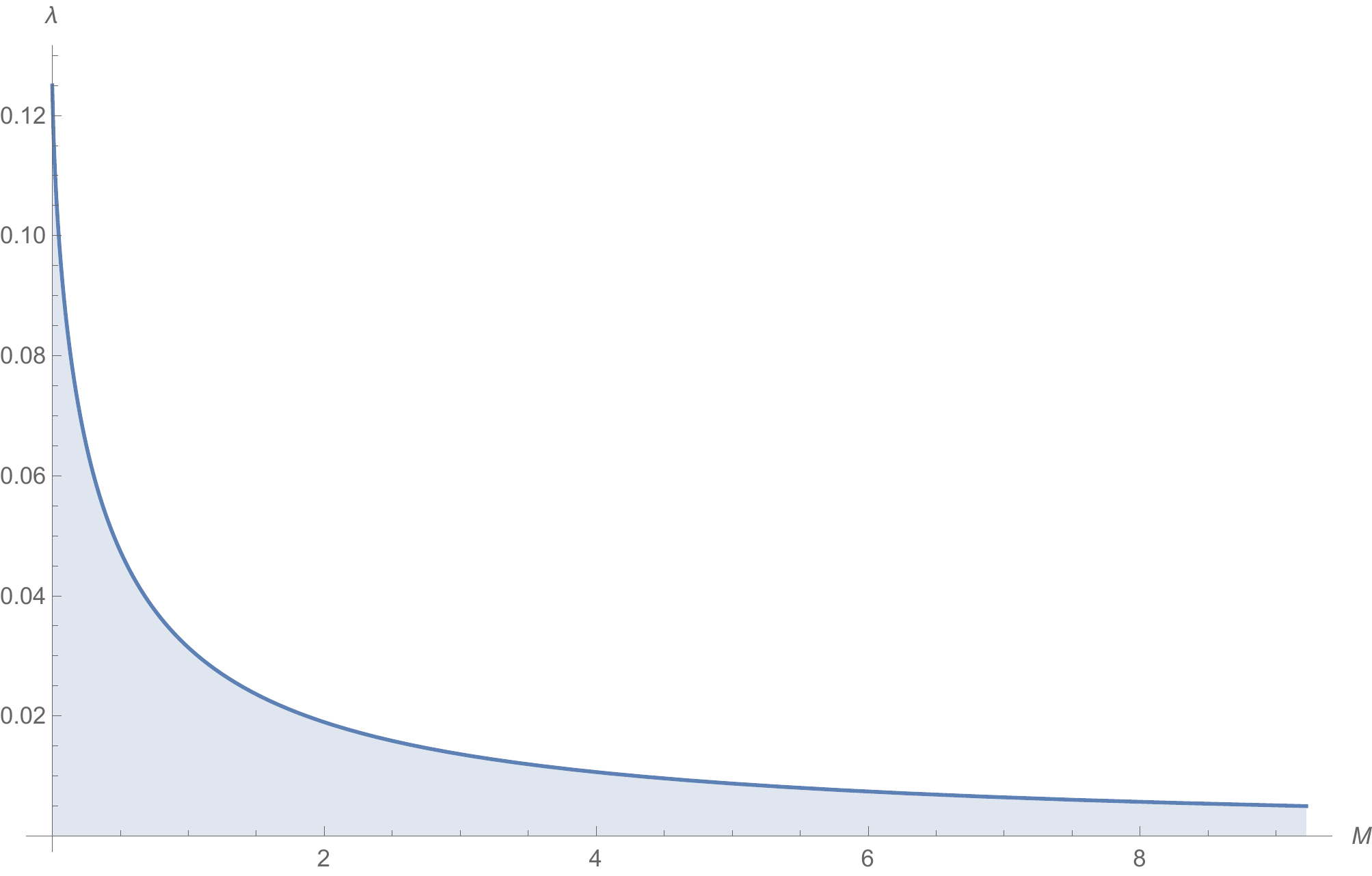}
    \caption{The critical curve $\lambda_c(M)$. The shaded area below the curve shows the allowed physical region on the $(M,\lambda)$ plane  where the density is real and positive.}
    \label{fig:critc}
\end{figure}

From our explicit result for the density \eq{densityreg} we find that imposing $\rho'(a)=0$ gives
\beq
\label{critline}
-\pi(1-p)py^3 -  9 Mp(K-E) y^2 + 9 M(-pK+  E)=0 \ .
\eeq
Combining this with \eq{x0res}, \eq{x1res} have three conditions linking the four variables $M,\lambda,p,y$, and thus we get a line on the $(M,\lambda)$ plane that we will call the critical line. We denote it by $\lambda_{c}(M)$ and we give a plot of it on figure \ref{fig:critc}. 

In order to study the critical regime analytically we can take the resultant in $y$ of \eq{x0res} and \eq{critline} which gives a 4th order polynomial in $M$ equation involving only $M$ and $p$, whose solution\footnote{note that we also have to pick the correct branch} is $M_{c}(p)$ which can be written explicitly but is rather lengthy. Furthermore, we can solve \eq{x0res} as a linear equation for $M$, then plugging the result into \eq{x1res} and \eq{critline} and taking their resultant in $y$ gives a 4th order polynomial equation whose solution is $\lambda_{c}(p)$. Thus we have a parametric representation of the critical line $\lambda_{c}(p)$ in terms of two (explicit but very lengthy) functions $M_{c}(p)$ and $\lambda_{c}(p)$.

\subsection{Merging of solutions and the physical region}

Numerically we observe that when $\lambda<\lambda_{c}(M)$ our two constraint equations \eq{x0res}, \eq{x1res} have two real solutions for $p,y$, with one of them being the actual physical solution while the other one should be discarded as it corresponds to a  non-positive density. As we move close to the critical line, we find that these solutions get closer and finally merge at the value $\lambda=\lambda_{c}(M)$. Beyond that point, i.e. for $\lambda>\lambda_{c}(M)$,  there is no solution with real positive density\footnote{There could be a two-cut solution in that region, whose exploration we postpone to the future}. Note that naturally the origin $(\lambda,M)=(0,0)$ lies in the allowed region since $\lambda$ and $M$ are weights in our partition function which is well defined for small enough values of them.

As a technical consistency check, let us show how
to derive $M_{c}(p)$ independently starting from our two original constraints \eq{x0res} and \eq{x1res}. For that we consider the equation \eq{cmM} which follows from them and reads $P(\lambda,M,p)=0$. Considering its lhs as a function of $p$ at fixed $\lambda,M$, we find that generically it has three roots $0<p_1<p_2<p_3<1$ of which $p_2$ is the physical one. It merges with $p_1$ at some $\lambda=\lambda(M)$ which will be the critical value and is characterized by the condition that  $\d_pP(\lambda,M,p)=0$.  Since this condition and the original equation $P(\lambda,M,p)=0$ are both polynomials in $\lambda$, we can take their resultant and find a polynomial equation now for $M(p)$, which is solved by the same function $M_{c}(p)$ we found using the shortcut $\rho'(a)=0$, i.e \eq{critline}. This shows that indeed  \eq{critline} corresponds to the boundary of the allowed parameter region, as expected. Notice also that the condition $\d_pP(\lambda,M,p)=0$ we just discussed means that on the critical line
\beq
    \d_p\lambda(M,p)=0
\eeq
which is\footnote{Notice that here we first differentiate $\lambda$ as a function of $M$ and $p$, and only then set $M=M_c(p)$} another equivalent formulation of the criticality condition.

\subsection{Looking for further singularities}

It could happen that on the critical curve itself there are special points at which additional phase transitions take place and the structure of the solution changes. So far we have not found any features of this type. For example, since $M_c(p)$ is given by a solution to a polynomial equation, one source of singularities could be the crossing of its roots at particular values of $p$. Indeed we can identify several such values by looking at the discriminant of this equation and numerically they are $p\simeq 0.05, \ p\simeq 0.29$. We find numerically that they do not seem to  correspond to a singularity of the physical solution but rather the crossing of singularities of unphysical roots. Similarly, for $\lambda_c(p)$ the discriminant of the equation that it solves vanishes at $p\simeq 0.09$ but again this does not seem to give a singularity in the physical solution.

\begin{figure}
    \centering
    \includegraphics[scale=0.5]{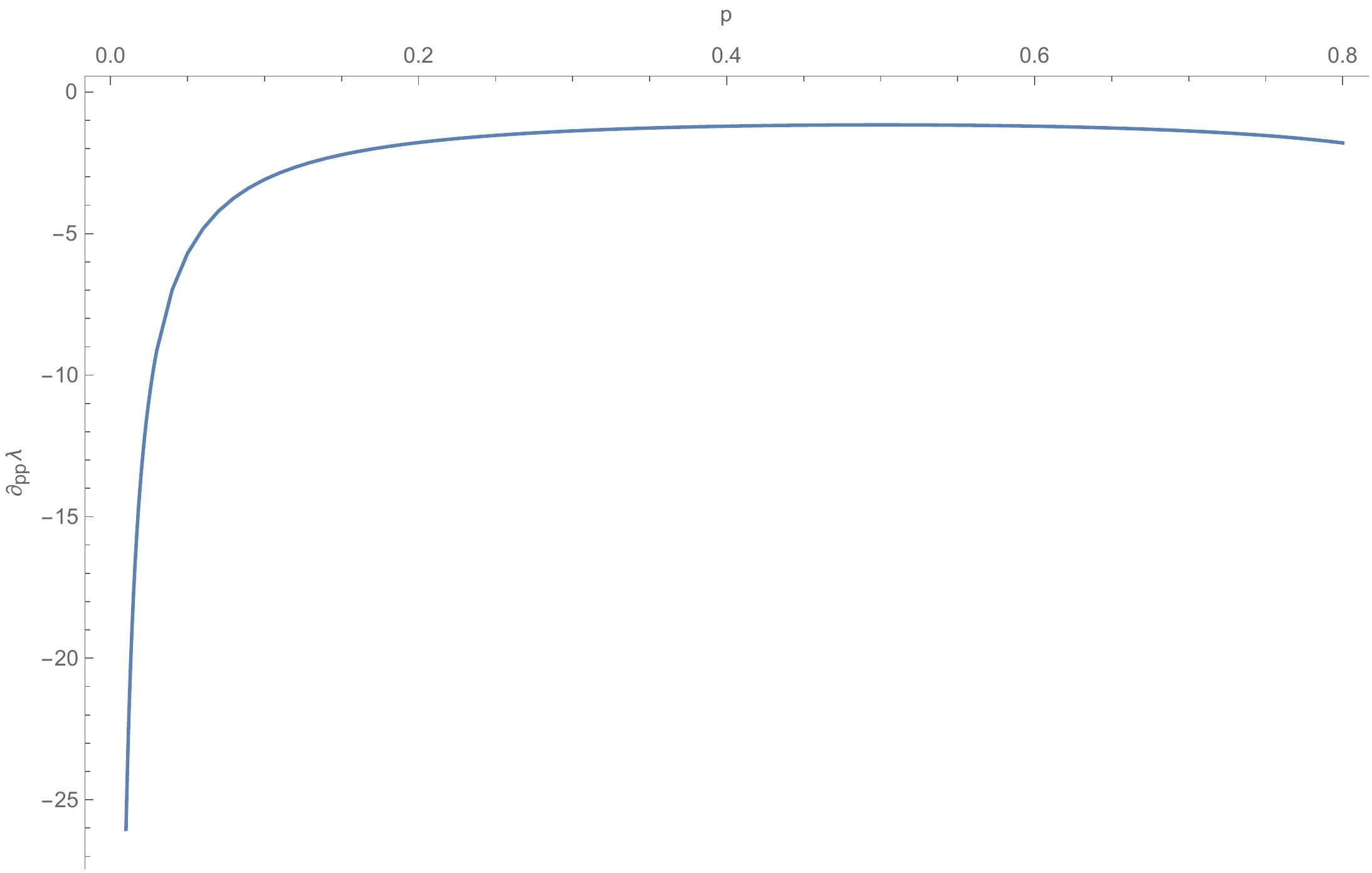}
    \caption{The derivative $\d_{pp}\lambda(M,p)$ evaluated on the critical line, shown as a function of $p$. We see no zeros or singularities for $0<p<1$.}
    \label{fig:d2}
\end{figure}

To further make sure that we are not missing any singularities, we can plot the 2nd derivative $\d_{pp}\lambda(p,M)$ at fixed $M$, evaluated on the critical line $M=M_c(p)$ (recall that the first derivative $\d_p\lambda(p,M)$ vanishes on the critical line). We give its plot on figure \ref{fig:d2}.  We find that it does not have any zeros or singularities and thus we would not expect to find any special points on the critical curve. A``physical" picture of eigenvalues behaving as Coulomb charges confined in a  potential well also does not suggest the existence of any additional phase transition for finite $\lambda$  and $M$ along the critical line:  the appropriate changes in these parameters only smoothly change the density of eigenvalues. The critical line corresponds to the values of $M_c(\lambda)$ when the eighevalues start spilling over the top of the potential~(see again Fig.\ref{fig:pot}).   We leave a more careful analysis for the future. 

\subsection{Decoupling of the heavy matter: $M \rightarrow \infty$ limit}

Let us discuss the limit of large mass when we expect the fermions to decouple so that we are left with the pure gravity limit. We see from  the matrix model potential that when
$\lambda \rightarrow 0$ the expansion of the square root yields the 
cubic term in the potential required for pure gravity.
To get finite coupling in pure gravity we have to take a  scaling limit 
$\lambda \rightarrow 0$, $M \rightarrow \infty$
with $M\lambda= \lambda_{\text{eff}}$  kept finite:
\begin{equation}
    V(z) = \dfrac{1}{2}z^2 -\dfrac{3}{2}\lambda_{\text{eff}} z^3 + O(\lambda^2) \ .
\end{equation}
Let us identify this limit in our critical curve. The regime $M \rightarrow \infty$, $\lambda \rightarrow 0$ corresponds to $p\to 1$ and 
 from the numerical solution for the critical line we see that the scaling is $M \sim \dfrac{1}{1-p} , \, \lambda \sim {1-p}, \ y \sim {1}$. Then from equations \eqref{x0res}, \eqref{x1res}, \eqref{critline} 
we find 
\begin{equation}
\begin{split}
        \lambda_c&=\frac{1-p}{16 \sqrt[4]{3}}+\frac{\left(31 \sqrt{3}-18\right) (1-p)^2}{768\ 3^{3/4}}+\dots \ ,
    \\ M_c&= \frac{16}{9\sqrt{3}}\frac{1}{1-p} + \left( -\frac{4}{9}-\frac{28}{27 \sqrt{3}} \right)+\dots \ .
\end{split}
\end{equation}
Therefore the critical value for the effective coupling gives:
\begin{equation}
   \left( \dfrac{3\lambda_{\mathrm{eff},\mathrm{crit}} }{2} \right)^2 = \dfrac{1}{108 \sqrt{3}} 
\end{equation}
which is in perfect agreement with the well known results from  \cite{Kazakov:1985ea} and \cite{Brezin:1977sv}. This is a nontrivial test of our calculation.

The emergence of the proper scaling parameter can be seen  in the
initial combinatorial partition function as well 
\begin{equation}
Z(\lambda,m)=\sum_{G}\lambda^{|G|}   \det[m^2+\Delta(G)] = 
\sum_{G}\lambda^{|G|} \sum_{F=(F_1\dots F_l)\in G}  \prod_{i=1}^{l} m^2 V(F_i) \ .
\end{equation}
Consider  the $m \rightarrow \infty $ limit (we recall that $M=2/9m^2$) and assume that the contribution 
from the term when all trees become the isolated nodes dominates. In this case 
$V(F_i)=1$ and we immediately get the $(\lambda m^2)^{|G|}$ weight factor which means
the logarithmic renormalization of the bare cosmological constant $\Lambda=-\log\lambda$ obtained from 
the matrix model.

\subsection{Critical line at \(M\to 0\)}

Having discussed the large $M$ limit in the previous subsection, in the remaining part of the paper we will mostly focus on the opposite limit of small $M$. Since $M$ controls the number of trees, for $M=0$ we have only 1 tree in the forest and thus for each graph we are counting the number of spanning trees on it with $\lambda$ weighing the number of graph's vertices. This is a well studied problem, describing in the critical regime the $2d$  QG in the presence of $c=-2$ matter, discussed in \cite{Kazakov:1985ea,Boulatov:1986jd,David:1985et,Kostov:1987kt,Klebanov:1990ip,Klebanov:1990sn,Edwards:1991jx} and the critical value is known to be $\lambda_c=1/8$. Indeed we verified  this is what we find from our critical curve (as can be seen already on the plot in figure \ref{fig:critc}). This is another successful test of our calculation. In subsequent parts of this paper we will make further contact with known results in this regime. 

 Let us also present here the expansion of the critical line near this point, which will be important for our further calculations. For the critical line $M\to 0$ corresponds to $p\to 0$ and it is convenient to use $p$ as an expansion parameter. We will also use instead of $\lambda$ a redefined coupling
\beq
    g=256\pi \lambda^2 \ .
\eeq
Then it is straightforward to obtain the expansion of the critical values $M_c(p)$ and $g_c(p)$. For $M_c(p)$ we have
\begin{align}\label{Mc(p)}
M_c(p)=A p +[B+C \log( p)]\,p^2+[D+E \log( p)+F \log^2( p)]\,p^3+\mathcal{O}\left(p^4\log^3( p)\right)\, 
\end{align}
where the coefficients read
\begin{align}
\label{Mcpv}
&A=\frac{2 \sqrt{2} \pi }{9}\,,\qquad  B=\frac{1}{18} \pi  \left(24 \pi -\sqrt{2} (41+\log (16))\right)\,,\qquad  C=\frac{\pi  }{9 \sqrt{2}}\,,\\ \nn &D=\frac{1}{288} \pi  \left(64 \pi  \left(21 \pi  \sqrt{2}-155+\log (16)\right)+\sqrt{2} (8385+8 \log (2) (173-168 \log (2)))\right)\,,\\ \nn &E=-\frac{1}{144} \pi  \left(32 \pi +\sqrt{2} (173-336 \log (2))\right)\,,\qquad  F=-\frac{7 \pi  }{12 \sqrt{2}}\,.
\end{align}
We see that each power of $p$ is accompanied by an expansion of a growing number of powers of $\log p$.
For the $g$ coupling we find
\begin{align}\label{gc(p)}
g_c(p)=A\,   +[B+C \log( p)]\,p+[D+E \log( p)+F \log^2( p)]\,p^2+\mathcal{O}\left(p^3\log^3( p)\right)\, 
\end{align}
with
\begin{align}
&A=4\pi\,,\qquad  B=\frac{16}{3} \left(3 \pi ^2 \sqrt{2}-\pi -12 \pi  \log (2)\right),\qquad  C=16\pi\,,\\ \nn &D=\frac{4}{3} \pi  \left(96 \pi ^2-171 \sqrt{2} \pi +38+96 \log ^2(2)-204 \sqrt{2} \pi  \log (2)+568 \log (2)\right)\,,\\ \nn &E=\frac{4}{3} \pi  \left(51 \pi  \sqrt{2}-142-48 \log (2)\right),\qquad  F=8\pi\,. \end{align}
Notice that the expansion starts with $g=4\pi+\dots$, corresponding of course to the critical value $\lambda_c=1/8$ discussed above.

We can also invert these expansions and find  $g_c(M)$ which reads to leading order
\beq
    g_c=4 \pi +\left(36 \sqrt{2} L_M+72 \pi -12 \sqrt{2}\right) M+O(M^2)
\eeq
where
\beq
    L_M=\log\frac{9M}{32\sqrt{2}\pi} \ .
\eeq
This gives the shape of the critical curve near its $M=0$ endpoint.

\section{New double scaling limit for 1-pt functions}
\label{sec:1pt}

At a generic point on the critical curve we expect the continuum theory to be pure gravity. However, when $M$ is strictly zero we have the very different $c=-2$ theory. This suggests to explore a double scaling limit when we get close to the critical line, in the vicinity of its $M=0$ endpoint. In this section we will define and study this near-critical regime in which we expect to see a nontrivial interpolation (flow) between the $c=0$ and $c=-2$ theories.

In order to design an interesting scaling limit, let us look at the 1-pt function $\langle\frac{1}{N}\tr\;\phi\rangle$ that we computed in closed form in \eq{phii1}. Technically it is convenient to use $p$ instead of $\lambda$ as an expansion parameter, as otherwise we would need to deal with iterated $\log \log$ corrections. The endpoint of the critical line corresponds to $(M,p)=(0,0)$, and we will consider an expansion near it with both $M$ and $p$ being  small. As a first example, the expansion of $\lambda$ or equivalently $g$ to the first few orders has the form 
\begin{align}
\label{gexp1}
    g&=
    4 \pi-16 \pi  p+32 \pi  p^2+O(p^3)
    \\ \nn
    &+M \left[\frac{\left(576 \left(\sqrt{2}-\pi \right) \pi -54 \sqrt{2} \pi  L+O(L^2)\right) p}{2 \pi }
    \right. \\ 
    &\ \ \ \ \ \ \ \ +\left.\frac{48 \pi  \sqrt{2}+144 \pi ^2+72 \pi  \sqrt{2} L+O(L^2)}{2 \pi }+O(p^2)\right]+ O(M^2)
\end{align}
with
\beq
    L=\log\frac{p}{16} \ .
\eeq
Thus at each order in $M$ we have a series in positive powers of $p$ and $\log p$. As a technical intermediate result, we also give the expansions of $a,b,c$ to higher order in appendix \ref{sec:smallMp}.

Next, expanding \eq{phii1} we find for the 1-pt function $\frac{1}{N}\langle\tr\;\phi\rangle$ 
\beqa
\label{phie1}
&&   \frac{1}{N}\langle\tr\;\phi\rangle=  
   4-\frac{128 \sqrt{2}}{15 \pi }+\left(8-\frac{96 \sqrt{2}}{5 \pi }\right) p\\ \nn
    &&
  +M \left[
     \frac{12 \left(124+11 \sqrt{2} \pi -15 \pi ^2\right)}{5 \pi
   ^2}+\frac{\left(432-90 \sqrt{2} \pi \right) L}{5 \pi ^2}
      \right.
      \\ \nn &&
      \left.     
      \ \ \ \  \ \ \ \ +p\left(-\frac{24 \left(3 \pi  \left(9 \sqrt{2}+5 \pi \right)-464\right)}{5 \pi
   ^2}+\frac{\left(6192-945 \sqrt{2} \pi \right) L}{10 \pi ^2}\right) 
   \right]+\dots
\eeqa
where like in \eq{gexp1} we dropped various higher order terms.

We see in both expansions \eq{gexp1} and \eq{phie1} a similar structure of a double series in $p$ and $M$ (with additional $\log p$ terms). This suggests to consider the limit when $p$ and $M$ both go to zero with their ratio fixed. Thus we define
\beq
\label{zdef}
   z=\frac{p}{16M}=\text{finite} \ , \ \ \ \ \ M\sim p\to 0
\eeq
and eliminate $p$ in favor of $z$. This gives a series now only in $M$, in which any given order receives contributions only from a finite number of therms in the original expansion and is thus straightforward to compute. Geometrically in this limit the branch points at $a$ and $c$ collide as they both approach $2$ with their difference being of order $p$, while the $[b,a]$ cut remains of a finite size as $b\simeq -2$.

Below we will compute in this limit the 1-pt functions $\frac{1}{N}\langle\tr\; \phi\rangle$ and $\frac{1}{N}\langle\tr\; \phi^3\rangle$. We will find that the results for them are closely related.

\subsection{The $\frac{1}{N}\langle\tr\; \phi\rangle$ one-point function}

Let us first discuss the beahavior of the 1pt function $\frac{1}{N}\langle\tr\; \phi\rangle$ in this scaling limit. We will need the result for the 1pt function to quadratic order only, and it reads
\begin{align}
\label{phie2}
\frac{1}{N}\langle\tr\; \phi\rangle&=-\frac{4 \left(32 \sqrt{2}-15 \pi \right)}{15 \pi }\\ \nn
&+ \frac{2 M}{5 \pi ^2} \left((-45 \sqrt{2} \pi  +216) \log (M z)+320 \pi ^2 z-768 \sqrt{2} \pi  z-90 \pi ^2+66 \pi  \sqrt{2}+744\right)+\\ \nn
&+\frac{M^2}{20 \pi ^3} \left[-20480 \sqrt{2} \pi ^2 z^2 \log (M z)-13608 \sqrt{2} \log ^2(M z)+4455 \pi  \log ^2(M z)\right.
\\ \nn &-30240 \sqrt{2} \pi ^2 z \log (M z)+198144 \pi  z \log (M z)-73008 \sqrt{2} \log (M z)
\\ \nn &-23004 \pi  \log (M z)+8100 \pi ^2 \sqrt{2} \log (M z)+40960 \pi ^3 z^2-182272 \sqrt{2} \pi ^2 z^2
\\ \nn &\left.-23040 \pi ^3 z-41472 \sqrt{2} \pi ^2 z+712704 \pi  z \right.
\\ \nn & \left.
+6480 \pi ^3+4536 \pi ^2 \sqrt{2}-74880 \sqrt{2}-108432 \pi\right] \ .
 \end{align}
Let us also define instead of $g$ the rescaled and shifted cosmological constant 
\beq
\label{Jdef}
J=2\pi\frac{4 \pi -g}{M}-\left(648 \sqrt{2} \pi -648 \pi ^2\right) M+144 \pi ^2+48 \sqrt{2} \pi
\eeq
which in the limit  \(M\to 0\) reads 
\beqa\label{j(z)}
J&=&512 \pi ^2 z-72 \sqrt{2} \pi  \log (M z)+\notag\\
 &+&M \Big[-81 \log ^2(M z)+864 \pi  \sqrt{2} z \log (M z)-324( \sqrt{2} \pi-1)  \log (M z)\notag\\
 && \ \ \ \ \ \ \ -16384 \pi ^2 z^2+9216\pi ( \pi-   \sqrt{2} )z\Big] \ .
\eeqa
The first term $2\pi\frac{4 \pi -g}{M}$ in the r.h.s. of \eq{Jdef} is a finite nontrivial quantity in the limit we consider (as follows from \eq{gexp1}) and can be viewed as a renormalized coupling in our regime. The other terms in that equation serve to remove from \eq{j(z)} the trivial nonsingular pieces that are polynomial in $M$ and do not contain $\log(M)$ or $z$, and thus will not affect the part of the 1-pt function we are interested in.

Similarly, subtracting from \eq{phie2} the irrelevant regular terms we define its singular part as
\beqa
\frac{1}{N}\langle\tr\; \phi\rangle_{sing}&=&\langle\frac{1}{N}\tr \;\phi\rangle+\frac{4 \left(32 \sqrt{2}-15 \pi \right)}{15 \pi }-\frac{\left(5 \sqrt{2} \pi -24\right) M J}{20 \sqrt{2} \pi ^3}\\ \nn
 &-& \frac{\left(1134 \pi ^2 \sqrt{2}-18720 \sqrt{2}+1620 \pi ^3-27108 \pi \right) M^2}{5 \pi ^3}
 \\ \nn
 &-&\frac{12 \left(11 \pi  \sqrt{2}-15 \pi ^2+124\right) M}{5 \pi ^2}
\eeqa
where the coefficient of the $MJ$ term is chosen so as to cancel the $M\log(Mz)$ term in \eq{phie2}. The final result reads
\begin{align}\label{phising}
 \langle\frac{1}{N}\tr \phi\rangle_{sing}&=\frac{1}{\pi^3}M^2\Big[ \left(486 \pi ^2 \sqrt{2}-3456 \sqrt{2}-1620 \pi \right) \log (M z)-243 \left(3 \sqrt{2}-\pi \right) \log ^2(M z)+\notag\\
 &+z \left( \left(10944 \pi -1728 \sqrt{2} \pi ^2\right) \log (M z)+ \left(5760 \pi ^2 \sqrt{2}-3456 \pi ^3+24576 \pi \right)\right)+\notag\\ 
&+z^2 \left( \left(6144 \pi ^3-18944 \sqrt{2} \pi ^2\right)-1024 \sqrt{2} \pi ^2 \log (M z)\right)\Big] \ .
\end{align}
Notice also that we can now retain in the definition \eqref{j(z)} of $J$  only the leading terms since it is already given by 2nd order in \(M\), and we denote the resulting quantity by little \(j\),
\begin{align}\label{j1(z)1}
j=512 \pi ^2 z-72 \sqrt{2} \pi  \log (M z) \ .
\end{align} 

To summarise, the singular part of the 1-pt function is given by \eqref{phising} where $z$ is implicitly a function of the coupling $g$ determined via the intermediate variable $j$ in \eq{j1(z)1} (or equivalently $J$), related to the coupling via \eq{Jdef}. These equations are one of our main results and
give the novel 1pt function representing the flow between \(c=-2 \)  and \(c=0\) regimes. 

Notice that according to \eq{Jdef} the variable $j$ is essentially the coupling $g$, up to constant factors, a shift and a rescaling by $M$. To write the 1-pt function in terms of $M$ and $g$ one would need to invert \eq{j1(z)1} to obtain $z(j)$. Curiously, this inverse function is (up to numerical constants that can be scaled away) in fact the well known Lambert function which has a variety of interpretations from combinatorics to quantum field theory and which appears here in our new result for the 1-pt function. We discuss its role and origins in more detail in section \ref{sec:lambert}.

\subsubsection{Limiting regimes}
\label{sec:1ptlim}

Let us now discuss the two limiting cases in more detail. In the $c=-2$ regime describing spanning trees we have to send \(z\to\infty\) as we take $M$ to be small and \(4 \pi -g\gg  M \). This gives \(z\simeq 2\pi\frac{4 \pi -g}{512 \pi ^2M}\). Plugging it into  \eqref{phising}  and retaining there the leading term  (last line) we get the   correct \(c=-2\) scaling
\begin{align}\label{c=-2}
 \langle\frac{1}{N}\tr\; \phi\rangle_{sing}&\simeq-\frac{1}{32\sqrt{2}\pi^3}(4 \pi -g)^2\log(4 \pi -g)\ .\end{align}
The dependence on $4\pi-g$, which is the parameter that measures deviation from criticality, is in complete agreement with predictions from \cite{Kazakov:1985ea,Boulatov:1986jd}.

In the latter case \(c=0\) pure gravity regime we have to solve the equation  \eqref{j1(z)1}
up to the 2nd order expansion around the critical\footnote{we recall that the critical line corresponds to $\d g(M,z)/\d z=0$} point  \(j'(z_c)=0\). We find 
\beq
\label{z0}
z_c=\frac{9}{32 \sqrt{2} \pi }
\eeq
which can be equivalently read off from \eq{Mc(p)}, \eq{Mcpv}. Taking $z$ to be near this critical value so that 
\beq
\label{zgcr}
z=\frac{9 }{32\sqrt{2} \pi }+\sqrt{\epsilon } \ , \ \ \text{where} \ \ \epsilon={\rm const\times} (g_c(M)-g)
\eeq
we  find the  scaling for the one-point function 
\begin{align}\label{c=0}
 \langle\frac{1}{N}\tr \;\phi\rangle_{sing}&\sim  \epsilon^{3/2}.\end{align}
Remarkably, the term \(  \epsilon^{1/2}\)  cancels, as it should be! This behavior thus perfectly matches the prediction from \cite{David:1984tx,Kazakov:1985ds}.

\subsubsection{Rational representation and rescaled form}

Let us present some other useful representations of the singular part of the 1-pt function. One natural way to rewrite it is to exclude  \(\log(z M)\)  from  \eqref{phising}  using \eqref{j1(z)1}. Then we obtain
\begin{align}\label{Phirational}
\Phi_{s}\equiv 1152\pi^5\frac{1}{M^2}\frac{1}{N}\langle\tr\; \phi\rangle_{sing}=16384 \pi ^3 j z^2-4608 \pi ^2 \sqrt{2} j z-8388608 \pi ^5 z^3+1769472 \pi ^4 \sqrt{2} z^2
\end{align}
where we dropped the (polynomial in $j$) terms that do not contain $z$ and thus do not affect the critical behavior. We see that in terms of $z$ and $j$ the result is purely polynomial. This form can be used just as the one above for analysing the asymptotic regimes. For example, to find the \(c=-2\) behavior we solve \eq{j1(z)1} iteratively and find
\begin{align}\label{j1(z)2}
 z\simeq\frac{j+72 \sqrt{2} \pi  \log (M \frac{j}{512 \pi ^2})}{512 \pi ^2} \ .
\end{align}
Then we plug this into
\eqref{Phirational} and take the leading term, which is precisely the result
\eqref{c=-2} we had before. The \(c=0\)  limit \eqref{c=0} is also straightforward to take.

\begin{figure}
\centering
\includegraphics[scale=0.3]{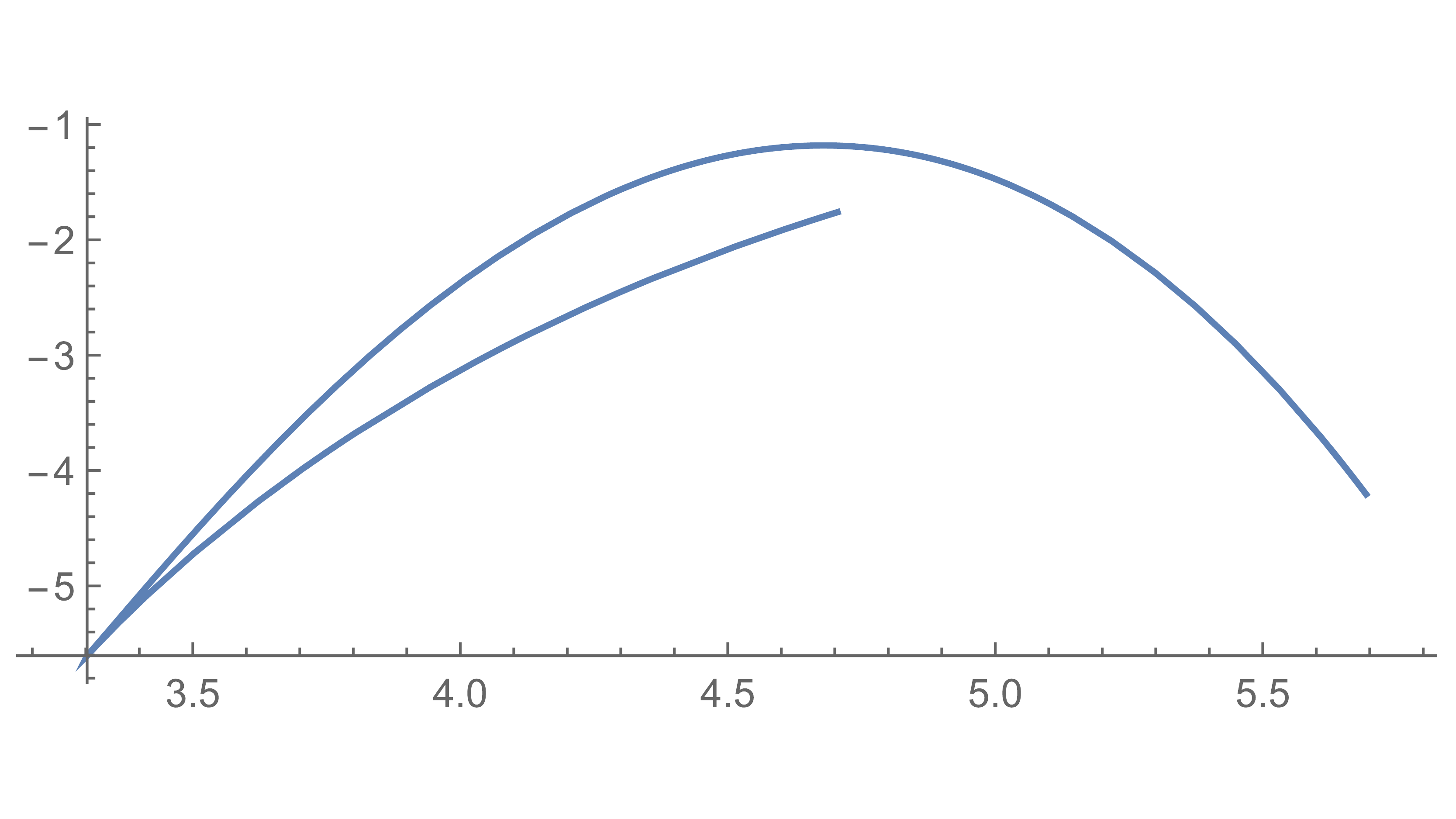}
\caption{Plot of singular part of \(\langle\frac{1}{N}\tr \phi\rangle\) given by \eq{Phiscaled} (without the overall constant $C$) as a function of  normalized coupling,  \({\cal J}={\Delta}/{\mu}\)  for fixed \(\mu=0.1\). At  \({\cal J}\simeq 3.30259\)  we see the \(c=0\)  singularity \(\langle\frac{1}{N}\tr \phi\rangle\sim (g_c^{(c=0)}-g)^{3/2}\).  The \(c=-2\) singularity \(\langle\frac{1}{N}\tr \phi\rangle\sim (g_c^{(c=-2)}-g)^{2}\log(g_c^{(c=-2)}-g)\)  occurs at  \(\mu\to 0\)  at fixed \(g\), on the 1st sheet of the function (upper branch in the picture).  The  exponential singularity is on the 2nd sheet (w.r.t. the \(c=0 \) branchpoint, lower branch on the picture).       }
\label{fig:plotGPhi}\end{figure}

Another useful rewriting can be done by absorbing various coefficients into new rescaled variables. Thus we redefine the variables in \eqref{j1(z)1}
 and \eqref{Phirational} as
 \begin{equation}
 \label{redefs}
z= \frac{t}{\frac{32 \sqrt{2} \pi }{9}}\,,\qquad M= \frac{32}{9} \sqrt{2} \pi  \mu\,,
\end{equation}
\beq
g= 4 \pi -256 \pi \Delta+24 \left(\sqrt{2}+3 \pi \right) M-324 \left(\sqrt{2}-\pi \right) M^2\,
\eeq
and lastly
\beq
    {\cal J}=\frac{\Delta }{\mu }=\frac{j}{72\sqrt{2}\pi} \ .
\eeq
Then the coupling  parameterization \eq{j1(z)1} becomes simply 
\begin{align}\label{J(t)}
{\cal J}=  t-  \log (\mu  t) \ .
\end{align}
 In these variables the critical value of $z=z_c$ corresponds to $t_c=1$. The physical range of ${\cal J}$ is then from $1-\log\mu$ to infinity (corresponding to $z_c<z<\infty$). For the singular part of our one-point function from \eq{Phirational} we get
\begin{align}\label{Phiscaled}
\Phi_s&= C\Big[2{\cal J}(  t^2-2  t)+3   t^2-2  t^3\Big]\,,
\end{align}
where \(C=23328 \pi ^2 \sqrt{2}\).
 Equations \eqref{Phiscaled}  and \eqref{J(t)}  represent the canonical parameterization of the (singular part of the) novel one point function \(\langle\frac{1}{N}\tr \;\phi\rangle\). This is one of the main results of this paper. We give a plot of the singular part of the 1-point function on figure \ref{fig:plotGPhi}.

\subsubsection{``Asymptotic freedom"  on the second sheet}

Let us consider the case \(j\to 0\). This is of course already beyond the \(c=0\)  criticality, which means that these are the ``unphysical"  values of parameters. However the underlying physical quantity may have  singularities on the second sheet which correspond to subleading exponential corrections.  Then we have   from  \eqref{phising} \begin{align}\label{assf}
z\simeq M^{-1}e^{-\frac{j  }{72 \sqrt{2} \pi\,  }}\,.
\end{align}
Plugging it into \eqref{Phirational} and picking the leading term we find a behavior reminding asymptotic freedom \begin{align}\label{Phirational2}
\Phi_{s}\simeq-4608 \pi ^2 \sqrt{2\,\,} (4 \pi -g)\,\,  e^{-\frac{4 \pi -g}{72 \sqrt{2} \pi\,M  }}\,,\qquad  \qquad (M\ll 4\pi-g) \ .
\end{align}

What are the excitations leading to these exponential effects? It cannot be the analogs of ZZ branes known for 2d gravity, since these ones have the \(e^{-{\rm const\times} N}\) behavior. They seem to be corrections to the regime of ``almost spanning trees", already for the leading order of planar graphs.  We leave a more detailed exploration and interpretation of this regime for the future.

\subsection{The $\frac{1}{N}\langle\tr\; \phi^3\rangle$ one-point function}

\begin{figure}
\centering
\includegraphics[scale=0.6]{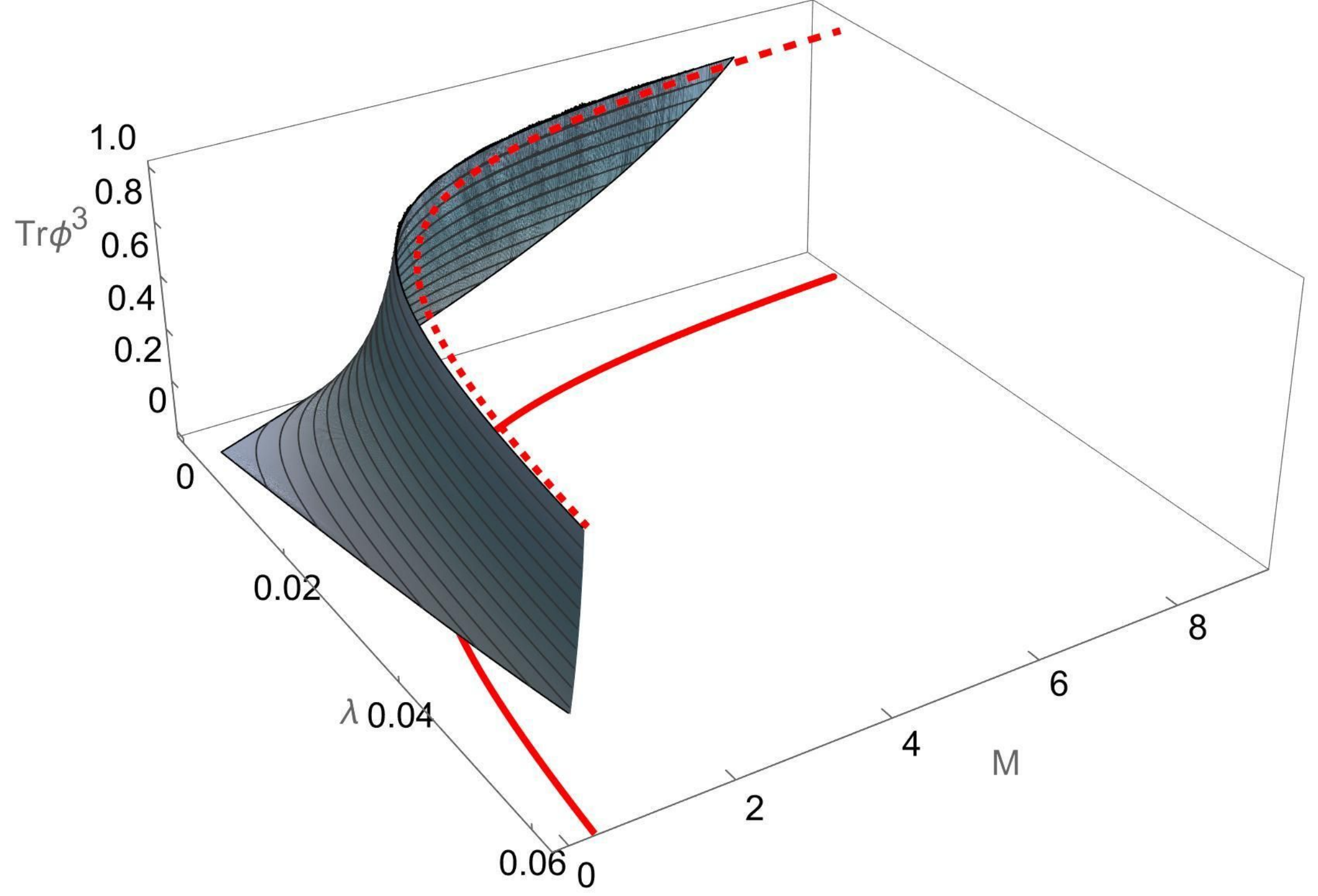}
\caption{Plot of the 1-pt function  $\frac{1}{N}\langle\tr\; \phi^3\rangle$ as a function of $M$ and $\lambda$, in the physical region bounded by the critical line. We show the critical line in red, in the planes $\frac{1}{N}\langle\tr\; \phi^3\rangle=1$ and $\frac{1}{N}\langle\tr\; \phi^3\rangle=0$.}
\label{fig:phi3}\end{figure}

Having studied above the 1-pt function $\frac{1}{N}\langle\tr\; \phi\rangle$, here we will discuss another one, namely $\frac{1}{N}\langle\tr\; \phi^3\rangle$. This 1-pt function is particularly important as it is related to the derivative of the partition function in the mass $M$ and consequently to the fermion condensate of the type $\langle \bar \psi \psi\rangle$. As discussed above in section \ref{sec:excorr}, it can be explicitly computed at generic values of the parameters and the result is given in the Mathematica file accompanying this paper. As an illustration,  figure \ref{fig:phi3} shows a 3d plot of this observable as a function of $M$ and $\lambda$ in the physical region. Here we will study its expansion in the scaling limit we just discussed above.

For this correlator we find expansions similar to what we had before, starting with the small $M$ expansion
\beqa
    \frac{1}{N}\langle \tr\;\phi^3\rangle&=&
    -\frac{59392 \sqrt{2}}{105 \pi }+256
    \\ \nn
     &+&\frac{64 M \left(-90 \left(21 \sqrt{2} \pi -94\right) \log (M z)+105 \pi ^2 (128 z-27)-8
   \sqrt{2} \pi  (3760 z-333)+16700\right)}{35 \pi ^2}
   \\ \nn 
    &-&\frac{32 M^2}{7 \pi ^3} \left(3 \left(7 \sqrt{2} \pi ^2 \left(512 z^2+5616 z-1215\right)-6 \pi  (30112 z-2997)+47562 \sqrt{2}\right) \log (M z)
    \right.
    \\ \nn &&
    \left.
    +243 \left(159
   \sqrt{2}-70 \pi \right) \log ^2(M z)+2 \left(-63 \pi ^3 \left(2048 z^2-864 z+135\right)
   \right.\right.
   \\ \nn &&
   \left.\left.
   +\sqrt{2} \pi ^2 \left(315520 z^2+24624 z-7047\right)+\pi 
   (96624-692800 z)+57624 \sqrt{2}\right)\right)
\eeqa
Then like before we define the singular part as
\beq
    \frac{1}{N}\langle \tr\;\phi^3\rangle_{sing}=\left[\left.\frac{1}{N}\langle \tr\;\phi^3\rangle-\frac{16 \left(21 \pi -47 \sqrt{2}\right)}{7 \pi ^3}MJ\right]\right|_{M^2}
\eeq
where we indicate that we take the term of order $M^2$ (dropping the terms of order $M^0$ and $M^1$ which like for $\frac{1}{N}\langle \tr\;\phi\rangle$ are regular). This gives 
\beqa \nn
\frac{1}{N}\langle \tr\;\phi^3\rangle_{sing}&=&
    -\frac{16 M^2}{7 \pi ^3} \left(42 \left(\sqrt{2} \pi ^2 \left(512 z^2+6048 z-1377\right)-96 \pi  (289 z-36)+6432 \sqrt{2}\right) \log (M z) \right.
    \\  &&
    \left.
    +567 \left(143
   \sqrt{2}-63 \pi \right) \log ^2(M z)+4 \left(-21 \pi ^3 \left(10240 z^2-4896 z+405\right)\right.\right.
   \\ \nn &&
   \left.\left.
   +9 \sqrt{2} \pi ^2 \left(56448 z^2-14672
   z-783\right)+\pi  (96624-476224 z)+57624 \sqrt{2}\right)\right)
\eeqa
Finally rewriting this in terms of ${\cal J}$ we find, up to an overall factor, the result analogous to \eq{Phiscaled} for this correlator,\footnote{here $\Phi_{3,s}$ is defined as $\frac{1}{N}\langle \tr\;\phi^3\rangle_{sing}$ multiplied by an overall constant and in which, like before, we further drop terms polynomial in $j$}
\beqa 
\label{Phi3s}
    \Phi_{3,s}&=&2 ({\cal J}-6\sqrt{2}\pi+8) (t^2-2t) + 3 t^2- 2 t^3 \ .
\eeqa
Notice that this function up to an overall multiplier coincides with the one for $\tr\;\phi$ defined in \eq{Phiscaled} up to an overall multiplier and redefinition ${\cal J}\to{\cal J}+{\rm const}$. This is a remarkable property which indicates a kind of universality for these 1-pt functions that would be important to elucidate further. As consequence, this function just like $\langle\tr\;\phi\rangle$ also interpolates perfectly between the $c=-2$ scaling \eq{c=-2} and $c=0$ scaling \eq{c=0}, with the limiting result having exactly the same form and the only difference being the overall constant factor. Furthermore, we see that both 1-pt functions are of course written in terms of the Lambert function which technically originates from the relation between ${\cal J}$ and the couplings in \eq{J(t)} which is the same for all observables of this type. 

In principle one should be able to compute further 1-pt functions of the type $\langle\tr\;\phi^n\rangle$ at least for fixed $n$ explicitly, and we leave this to future work.

\section{Disc partition function}
\label{sec:disc}

While in section \ref{sec:1pt} we have studied extensively the 1-pt function, here we will discuss a more complicated observable -- the disc partition function. We will explore its critical behavior and the interpolation (flow) between $c=-2$ and $c=0$ regimes.

We will study two resolvents: the resolvent $G(x)$ which generates $\langle\tr X^k\rangle$ correlators,
\beq
     G(x)=\sum_{k=0}^\infty \frac{1}{x^{k+1}}\frac{1}{N}\langle\tr\; X^k\rangle=\int_b^ady\frac{\rho(y)}{x-y}
\eeq
and another resolvent \beq
    H(r)=\sum_{k=0}^\infty \frac{1}{r^{k+1}}\frac{1}{N}\langle\tr\; \phi^k\rangle=\int_b^ady\frac{\rho(y)}{r-\phi(y)}
\eeq
generating the correlators $\langle\tr\;\phi^k\rangle$. These resolvents can also be viewed as disc partition functions, with $k$ in the correlator being the disc boundary length, corresponding to different boundary conditions (see section \ref{sec:bound} for details). We will show that they have the same universal behavior in the pure gravity limit as expected, while for the $c=-2$ regime we find different properties depending on the type of boundary.

\subsection{The resolvent $G(x)$ for $\tr\; X^k$ correlators}

First let us discuss the $G(x)$ resolvent. We will work in the same limit as in section \ref{sec:1pt} described in \eq{zdef}, so that $M\sim p\to 0$. We will be interested in the singularity of the resolvent near its branch point which physically describes the situation when the boundary of the disc becomes large. Thus we expand $G(x)$ around the branch point at $x=a$ (which in our limit collides with the $x=c$ branch point present on the other sheet). In our limit we have
\beq
\label{abcexp2}
    b=-2+O(M) \ , \ a\simeq 2+M\alpha \ , \  c\simeq a +64 Mz
\eeq
where
\beq
    \alpha=\frac{3(3\sqrt{2}\log(Mz)+6\pi+2\sqrt{2})}{\pi}
\eeq
and we expand $x$ as 
\beq
    x=2+M\chi  
\eeq
with $\chi \sim 1$ being the rescaled boundary cosmological constant. Then,  using \eq{piexpd} and dropping the terms which are regular in $\chi$,  we get
\beq\label{Gsing}
    G_{sing}=\sqrt{M}\left[\frac{-9\sqrt{2}\cosh^{-1}\left(\sqrt{\frac{\chi-\alpha}{64z}}\right)}{\pi\sqrt{\chi-\alpha-64z}}-\sqrt{\chi-\alpha}\right]+O(M)
\eeq
where we indicated that we keep only the singular terms. The singularity is at $\chi=\alpha$ (which in view of \eq{abcexp2} corresponds of course to $x\simeq a$) and comes about from the square root and the $\cosh^{-1}$ terms. Notice that the apparent singularity at $\chi=\alpha+64z$ in fact cancels. 

\subsubsection{Limiting cases}
\label{sec:Glim}

Equation \eq{Gsing} gives the singular part of the resolvent which describes the critical behavior. Let us examine its $c=0$ and $c=-2$ limits. In the former regime (pure gravity) we should expand as in section \ref{sec:1ptlim}  around the critical value of $z$ given by  $z_c=\frac{9}{32\sqrt{2}\pi}$ (see \eq{z0}), and the same time we expand in $\chi$ near the singularity,
\beq
    z=z_c+\delta \ , \ \ \chi=\alpha_c+64\delta/ \xi \ , \ \ \delta\to 0
\eeq
with $\alpha_c=(\left.\alpha)\right|_{z=z_c}$ and the factor $64$ introduced for convenience. The variable $\xi$ is a finite scaling parameter and $\delta/\xi$ corresponds to the rescaled boundary cosmological constant while $\delta$ is viewed as the square root of the renormalised bulk cosmological constant (due to the relation \eq{zgcr} between $z$ and the coupling in this limit). Then we get with $\delta\to 0$
\beq
    \frac{1}{\sqrt{M}}G_{sing}=c_1+c_2\frac{\delta}{\xi}+c_3\frac{\delta^{3/2}}{\xi^{3/2}}(\xi-2)\sqrt{\xi+1}+\cal{O}(\delta^2)
\eeq
where $c_n$ are (real) numerical constants. Thus we reproduce the universal pure gravity prediction of \cite{Kazakov:1989bc} -- the expected scaling function $(\xi-2)\sqrt{\xi+1}$ and the expected prefactor $(\delta/\xi)^{3/2}$. This is yet another important test of our results.

Next, the $c=-2$ regime corresponds to scaling $z\to \infty$ and we take at the same time $\chi\to\infty$ so that the ratio
\beq
    \kappa\equiv\frac{\chi}{z}
\eeq
remains fixed. Then we find 
\beq
    \frac{1}{\sqrt{M}}G_{sing}\simeq -\sqrt{\kappa}\sqrt{z} -\frac{9\sqrt{2}\cosh^{-1}\left(\sqrt{\frac{\kappa}{64}}\right)}{\pi\sqrt{(\kappa-64)z}} \ .
\eeq
The interpretation of this result remains to be clarified since the observables $\tr \;X^n$ are not very natural for the $c=-2$ limit. Below we will see however that for the other resolvent $H(r)$ corresponding to $\tr\;\phi^n$ correlators we recover perfectly the known results for the $c=-2$ limit.

\subsection{The resolvent $H(r)$ for $\tr\; \phi^k$ correlators}

While we have computed the resolvent $G(x)$ in closed form, the  resolvent $H(r)$ for $\tr\; \phi^k$ vevs is harder to obtain exactly and we leave this question for the future. Nevertheless we will be able to compute here the universal part of it which is responsible for critical behavior, i.e. the singular part.

As for $G(x)$, we will be interested in the singularity that appears when the argument approaches the branch point of the resolvent. For $H(r)$ the $[b,a]$ cut in $x$ gets mapped to the cut $[\phi(b),\phi(a)]$ in the $r$ variable, so we will focus on the regime $r\to \phi(a)$.  The most complicated part of the integral we wish to compute reads\footnote{Notice the $\Pi$ function here is regular on $[b,a]$ but has a $\sim 1/\sqrt{x-c}$ behavior at $x=c$.}
\beq
\label{intp1}
    \int_b^ady \frac{y-2c}{r/2-c+\sqrt{c}\sqrt{c-y}}\sqrt{\frac{a-y}{y-b}}\Pi\left(\frac{y-b}{c-b},1-p\right) \ .
\eeq
We will work as before in the limit $p\sim M\to0$ so $a$ and $c$ are close.  The singularity of the resolvent comes from the integration region $y\simeq a$ when the denominator becomes  close to zero since at the same time $r\simeq\phi(a)\simeq 2c$. To zoom in on this singularity we make the change of variables
\beq
    y=a-M Y \ .
\eeq
In order to get a nontrivial result we would like both terms $(r/2-c)$ and $\sqrt{c}\sqrt{c-y}$ in the denominator of \eq{intp1} to be of the same order (notice that both $y-a$ and $c-a$ are of order $M$), which means that $r-2c\sim \sqrt{M}$. Thus we define the rescaled variable $R\sim 1$ by
\beq
     r=2c+2\sqrt{M}R \ .
\eeq
Now let us discuss the behavior of the $\Pi$ function in \eq{intp1}. We see that its argument is $1-M\frac{Y+64 z}{4}$ and its modulus is $1-p=1-16Mz$, so both of them approach $1$ at the same rate $\propto M$. We did not find in the literature the expansion of $\Pi$ in this regime (which corresponds to a nontrivial resummation of the standard expansions in which one of the two arguments is held fixed) but it can be quite straightforwardly derived from the integral representation of $\Pi$.  The result reads
\beqa
\label{piexpd}
&&
    \Pi(1-S\epsilon,1-T\epsilon)\simeq \frac{1}{\epsilon}
    \frac{{\rm arccosh}(\sqrt{S/T})}{\sqrt{S(S-T)}} \ , \ \ \epsilon\to 0
\eeqa
which is also easy to verify numerically. Plugging in the values of our parameters and combining all the parts, for the full density \eq{densityreg} we find in our regime
\beq
    \rho\simeq \frac{\sqrt{M} \sqrt{Y }}{\pi }-\frac{9 \sqrt{2} \sqrt{M} \cosh ^{-1}\left(\frac{1}{8} \sqrt{\frac{Y }{z}+64}\right)}{\pi ^2 \sqrt{Y
   +64 z}} \ .
\eeq
Thus by focusing on the endpoint of the integration region we have got rid of elliptic functions. The integral in $Y$ can be now taken analytically as an indefinite integral. Plugging in the limits of integration, expanding for $M\to 0$ and discarding regular contributions we finally get for the singular part \beq
\label{Hsing}
    \frac{1}{M}H_{sing}(R)=\frac{\left(\pi  R\sqrt{2} \sqrt{128 z-R^2}-18 \arccos\left(\frac{R}{ \sqrt{128z} }\right)\right) \arccos\left(\frac{R}{\sqrt{128z}
}\right)}{\pi ^2} \ .
\eeq
We see that as expected it has nontrivial square root and logarithmic singularities in $R$ whose position moreover depends nontrivially on our finite scaling parameter $z$. The singularity is located as expected at the point where $r=\phi(a)$ which after expansion in $M$ translates to $R=-\sqrt{128z}$. It may seem that there is also a branch point at $R=+\sqrt{128z}$ but in fact it cancels between the different terms in \eq{Hsing}.

This singular part of the resolvent \eq{Hsing} is another one of our main results. Below we will discuss its limiting cases corresponding to the pure gravity and spanning trees regimes.

Remarkably, the analytic structure of \eqref{Hsing} reminds that of the analogous disc partition functions for the flattening of random geometries in the model of dually weighted graphs~\cite{Kazakov:1995ae,Kazakov:1995gm,Kazakov:1996zm,Kazakov:2021uio} (see also \cite{Kostov:1997bn}). This similarity deserves a further study which we postpone  to the future.

\subsubsection{Limiting cases}
To get the pure gravity limit, similarly to the discussion above in section \ref{sec:Glim}, we expand near the critical value of $z$. Thus we set
\beq
   z=z_c+\delta \ , \ \ R=-\sqrt{128z_c}+\frac{64\sqrt{\pi}}{3\;2^{3/4}}\frac{\delta}{X}
\eeq
where
\beq
    z_c=\frac{9}{32\sqrt{2}\pi}
\eeq
and expand for small $\delta$. This gives
\beq
    \frac{1}{M}H_{sing}^{(c=0)}\simeq c_1+c_2\frac{\delta(X+1)}{X}+c_3\frac{\delta^{3/2}}{X^{3/2}}(X-2)\sqrt{X+1}+\dots
\eeq
where $c_k$ are numerical constants. We see that the result perfectly reproduces the correct pure gravity scaling function \cite{Kazakov:1989bc} and the expected $3/2$ power of $\delta$ (notice there is no $\delta^{1/2}$ term).

Let us now consider the $c=-2$ limit. To do this we expand the scaled disc partition function  \eqref{Hsing} with \(z\simeq \frac{4 \pi -g}{256 \pi M}\sim R^2\to\infty\) (the result of  dropping the last term in \eqref{j1(z)1} in this limit). We thus keep finite the combination $\zeta=\frac{ R}{\sqrt{128z}}$ .
Then we find for the singular part of $H(r)$ 
\begin{align}\label{Hsingc=-2}
  H_{sing}^{(c=-2)}\simeq -\frac{ \sqrt{2}}{2\pi^2 } (4 \pi -g)  \zeta  \sqrt{\zeta ^2-1} \log \left(\sqrt{\zeta ^2-1}+\zeta \right) \ .
\end{align}
Remarkably, it perfectly coincides with the prediction for the $c=-2$ limit  obtained in \cite{Kostov:1992pn} (see equation (4.19) there) from a different matrix model. That shows once again the universality of the critical regime.

Thus we see that our result for the singular part of the resolvent interpolates nontrivially between two very different predictions in the $c=0$ and $c=-2$ regimes and describes the flow between these two models.

\section{Conclusion}
\label{conclusions}
\label{sec:conclusions}

In this study we investigated the model of massive spinless fermions interacting
with 2d quantum gravity. We derived the Hermitian matrix model with non-polynomial potential
describing the theory, and solved it in the planar approximation considering the one-cut solution. The regime where this solution exists is restricted
by a critical curve in the 2-dimensional parameter plane of fermion mass $m$ and cosmological coupling $\Lambda=g_c-g$.
It is explicitly demonstrated that the theory 
in the  scaling limit $m^2\sim\Lambda\to 0$ interpolates between
the $c=-2$ theory, for $m^2\ll \Lambda$, when the spanning trees dominate, and the pure 2d gravity $c=0$ theory, for $m^2\gg \Lambda$,
 when the fermions renormalize 
the cosmological constant in a simple way. We also computed the universal singular part of the disc 
partition functions in this scaling with Dirichlet and Neumann boundary conditions, interpolating between the $c=0$ and $c=-2$ regimes. They  fit perfectly with the previous results known for the limiting $c=0$. The former one also fits the known $c=-2$ regime whehter as the latter one demonstrates in this limit a new behavior.

According to the matrix-forest theorem for the 
massive determinant we  have identified the dominant number of trees in the forest in the
different regions at the parameter space. In other words we took into account 
the backreaction of the 
massive matter on the 2d quantum geometry. At $m=0$ the single tree saturates the partition function
at  criticality and we reproduce the picture for the $c=-2$ theory. 
The cosmological constant dependence of physical quantities (one-point functions, disc partition functions) contains logarithms due to the influence of "large" trees in this limit. At $m\rightarrow \infty$
the heavy matter breaks the 2d Euclidean space-time into the maximally possible number
of components which coincides with the number of zero modes of the graph Laplacian.

The most interesting behavior occurs at small but finite $m$ where we find a new scaling
behavior. The new scaling parameter is the ratio  $J\sim\frac{\Lambda}{m^2}$
and the parameterization  of the scaling functions  is given in terms of the Lambert function  \eqref{universalJ} of the parameter $t$. In this limit sufficiently large and numerous trees in the partition function matter and
it turns out that this scaling regime exists at a narrow region near the critical
curve. This scaling describes the critical flow between $c=-2$ (in UV) and $c=0$ (in IR) regimes.

There are a few questions  concerning our solution which would be interesting to clarify:
\begin{itemize}
\item  Study of the whole variety of flows in the vicinity of $c=0$  and $c=-2$  critical points. Comparison to another flow found in \cite{Kostov:2006ry}.  Generalization of such flows to all central charges of matter $c\le 1$. 
    \item Analysis of the rest of the parameter plane and of
     the multi-cut solutions to the matrix model.
     \item Exploring various 1-pt functions and understanding the origin of the simple relation between those we computed in \eq{Phiscaled} and \eq{Phi3s}
    \item Clarification of the role of the second solution to the quadratic equation 
     in the Parisi-Sourlas derivation of the matrix potential. 
   \item Computation of instanton contributions of different kinds, including ZZ branes.
   \item  Establishing the double scaling limit $N\to\infty, \Lambda\to 0$  of our model  along the critical line and deriving the universal scaling function in this limit.
   \item Derivation of this and other critical flows  from the continuous 2d QG (Liouville formalism).
\end{itemize}

We hope to return to these questions in our future research.

\section{Further directions}
\label{sec:further}

Here we outline in more detail some nontrivial potential directions for future exploration.

\subsection{Double scaling limit and sum over topologies}

It should be possible to solve our one-matrix model in the double-scaling limit~\cite{Brezin:1990rb,Douglas:1989dd,Gross:1989vs}, for the whole critical flow in the space of $(M,\lambda)$. It would be interesting to embed such a double scaling solution into the KdV formalism of~\cite{Douglas:1989ve} for 2d gravity interacting with $c<1$ matter fields.  We could also expect other integrability pattern in our double scaling limit with the Toda hierarchy involved like in \cite{nekrasov}. This is quite common for the theories with asymptotic freedom. The emergence of the Lambert function supports this expectation.

Is it possible to get the analogue of the Kontsevish matrix 
    model in our case? The fermionic bilinears (derivative with respect to mass)
    should reproduce some classes at the moduli space. In the double scaling limit
    the Lambert function emerges implying the  relation with the
    Hurwitz numbers which indeed according to ELSV formulae are written as 
    particular integrals over the moduli space.

\subsection{Generalization to other critical flows}

An obvious generalization of our  formulas describing the universal flow between $c=-2$ and $c=0$ critical points \eqref{universalFi}, \eqref{universalJ}, \eqref{Gsing}, \eqref{Hsing} would be the construction of a more general  model with forests on planar graphs, working for all central charges $-\infty <c<1$. Following the Kastelyn-Fortuin-Stephens tree expansion \cite{stephen1976percolation} for the $Q$-state Potts model, realized on planar graphs as a Q-matrix model  in~\cite{Kazakov:1987qg}, we have to introduce the loops into the trees and weigh such configuration with an extra factor $Q^{\# {\rm loops}}$. The central charge will depend on $Q$. It would be interesting to construct such a (multi)matrix model, which will be different from the model of \cite{Kazakov:1987qg} since we deal here with rooted trees. As we have seen on the example of the flow between  $c=-2$ and $c=0$ in our paper, such a model would have  universal critical flows different from those of the $O(n)$ model of Kostov~\cite{Kostov:1988fy}.    

Furthermore, in \cite{Bondesan:2016osa} an interesting mapping of the model of unrooted trees considered there  to the (loop+dimer) statistical model of  Kostov and Staudacher \cite{Kostov:1992pn} has been developed. What is the similar corresponding statistical
     model in our case of rooted trees?

\subsection{Fragmentation of the RRG into finite number of trees and many-body localization} 
Recently the RRG ensemble  has attracted a lot of attention being the toy model for a Hilbert space of
some interacting many-body problem (see \cite{mirlin} for review and references therein). One considers
the  spinless fermion $\psi_i, i=1\dots N$ on RRG at large $N$  with diagonal on-site
disorder. The partition function of the model reads
\beq
Z(W,N)=\sum_{RRG}d\psi d\psi^{\dagger} \exp(\psi^{\dagger}_i(L_{ij}+\delta_{ij}\epsilon_i)\psi_j)
\label{loc}
\eeq
where $\epsilon_i$ is the random diagonal disorder with the flat distribution 
$\epsilon_i\in (-W,W)$. Let us compare this 
model with our study. Both models describe the spinless massive fermion interacting with 
the 2d gravity, although in (\ref{loc}) the mass of the fermion is random while we
consider the fermion with the fixed mass.
The second difference is that we 
consider the canonical ensemble with the cosmological constant while in (\ref{loc})
the microcanonical ensemble is assumed. 

The models are very close but the questions discussed are quite different. 
We have integrated out the fermions and look at the emergent partition function
with the determinant in the critical regime of large number of nodes as a function of two couplings.
The usual question in the model (\ref{loc}) is different and concerns the localization 
or delocalization
of the fermion at the graph due to the disorder. It was found \cite{mirlin} that there exists a
critical $W_{cr}$ such that for $W>W_{cr}$ the Anderson localization takes place. This
has been established via numerical evaluation of level spacing distribution
or IPR.

The one-particle Anderson localization on RRG itself seems to be a somewhat artificial
problem, however it becomes  interesting if we treat the RRG ensemble as 
the model of Hilbert space for some interacting many-body system. The mapping is
not exact but it captures the key qualitative features. 
The one-particle Andersen localization transition
in the Hilbert space is treated according to the conjecture from \cite{levitov}
as the transition to the many-body localized (MBL) phase in the physical space. 
It was argued (see \cite{Moudgalya_2022} for review and references therein) that the 
fragmentation of the Hilbert space 
is one of the key mechanisms for the transition into the MBP phase. In terms
of RRG it means that we are looking at the strong backreaction of fermions leading to
fragmentation of RRG into some 
number of weakly connected subgraphs, possibly trees.

Another mechanism of Hilbert space fragmentation involves not the 
random fermion mass but the perturbation of the RRG ensemble by the
chemical potentials for the short cycles. 
These chemical potentials 
can be considered as the leading terms of the expansion 
of the characteristic polynomial of the graph Laplacian  in inverse powers of $m^2$ since \(\tr A^n\) 
gives the number of cycles of length \(n\) on the graph
$\log \det(A+m^2)\propto \sum_k (m^2)^{-k} \Tr A^k$. Here we 
have a clear-cut link to our model in the large $m^2$ limit.

If a cubic perturbation \(V=t_3\Tr A^3\) is chosen,   the phase transition occurs at
some critical value \(t_{3,crit}\) and the RRG ensemble  at \(t_3>t_{3,crit}\)
gets dominated by the
clustered graph with the number of clusters equal  to \(N_{cl}=\frac{|G|}{q}\) \cite{avetisov2016eigenvalue}.
Similar graph fragmentation occurs for \(V=t_4\Tr A^4\) at \(t_{4,crit}\) when
bipartite clusters  emerge \cite{kelly2019self,valba2021interacting}. From the spectral viewpoint
of the graph Laplacian each cluster represents the single low energy mode
escaped from continuum.  The isolated eigenvalues form the second
soft "non-perturbative" band in the spectrum \cite{avetisov2016eigenvalue}.
It turns out that the spectrum of the perturbed RRG ensemble enjoys the
mobility edge which separates the delocalized states in the main part of the spectrum 
and localized modes in the second non-perturbative band \cite{avetisov2020localization}.

We do not expand the massive determinant, hence the chemical potentials for
all cycles are present. The number of trees in the forest, which is the
order parameter for the Hilbert space fragmentation, in our exact
solution in the planar limit depends on the mass
and the cosmological constant 
which provides the soft cut-off in the
Hilbert space dimension. At small mass we have no Hilbert space fragmentation 
at all. The investigation of the localization 
of the fermions in the gravity background in our model is a clear 
direction for further study. In order to explore this, more detailed characteristics
like the level spacing distribution have to be analysed. We postpone 
this issue for a separate investigation.

We could try to solve this model in the double scaling limit, summing up over the topologies in the critical regime (near our critical curve, and presumably at $m\to 0$). One might hope (though it is not at all guaranteed) that the double scaling solution contains this fragmentation phenomenon.

\subsection{ZZ-brane instantons for one-cut solution}

Let us make a comment  concerning the instanton effects and
consider the single eigenvalue ZZ-brane instanton for our one-cut solution.
The instanton action evaluated along the spectral curve reads as follows
\begin{equation}
S_{inst}=\int_{a}^{x_0} Y(x)dx
\end{equation}
where the spectral curve in terms of the resolvent reads 
\begin{equation}
    Y(x)= V'_{eff}(x)= V'(x) - 2G(x)= M(x)\sqrt{(x-a)(x-b)} \ .
\end{equation}
The critical point $x_0$ of the effective potential $V_{eff}(x)$  is defined
by condition $M(x_0)=0$ and corresponds to the pinch point of the spectral curve.
The effective potential is constant on the cut and its derivative obeys the useful 
relation \cite{DiFrancesco:1993cyw}
\begin{equation}\label{veff}
  \partial _{\lambda}V_{eff}(a)= 2\log \frac{(a-b)}{4}  
\end{equation}

The instanton action can
be written as 
\begin{equation}
    S_{inst}= V_{eff}(a)-V_{eff}(x_0) \ .
\end{equation}
Let us focus on the instanton contribution at $M\rightarrow 0$
when we can approximate the rhs by derivative since 
$x_0$ is close to $a$,
\begin{equation}
    S_{inst}\propto \partial_a V_{eff}(a)(x_0-a)\propto \frac{\partial c}{\partial a} 
    \frac{\partial V_{eff}}{\partial c}(a)(x_0 -a)
    \propto \log (b-a)   (x_0-a) \ .
\end{equation}
We recall that $c= \frac{1}{4\lambda}$. Using the identity (\ref{veff}) and expansions for $a,c$ from appendix \ref{sec:smallMp} one can check that $\frac{\partial c}{\partial a}$ is finite in this limit. We see
that the instanton action vanishes at $M\rightarrow 0$ and instanton contributions 
become unsuppressed.

\subsection{Kesten-McKay distribution and  criticality without planarity}

So far we have considered the planar approximation but here we shall
make a short remark concerning the partition function for the generic RRG 
microcanonical ensemble at large number of 
nodes $n\rightarrow \infty$. 
In this limit we can 
utilize the famous Kesten-McKay(KM) distribution for the spectral 
density of RRG ensemble \cite{kesten1959symmetric,mckay1981expected}.

Consider the derivative of our partition function  $Z(m^2)$ at the critical line
which yields the resolvent of Laplacian for RRG $R_n(M)$
\beq
R_n(m^2) = \frac{1}{n} \langle \Tr \frac{1}{L-m^2}\rangle = -\frac{2}{n}\frac{d}{dm^2}\langle\log Z(m^2)\rangle
\eeq
The resolvent of the adjacency matrix of RRG reads \cite{kesten1959symmetric,mckay1981expected}
\beq
\label{resRRG}
R_n(z)= \frac{(2-q)z +q\sqrt{z^2- 4(q-1)}}{2(z^2 -q^2)} \ 
\eeq
and the resolvent of the Laplacian of RRG can
be derived via the shift $z\rightarrow q-z$.

Hence from the spectral density of RRG
\beq
\rho_n(\lambda,q)=\frac{1}{n}\langle\sum \delta(\lambda -\lambda_i)\rangle_{RRG}=
\frac{q}{2\pi} \frac{\sqrt{4(q-1) -\lambda^2}}{q^2 -\lambda^2} + O(1/n)
\eeq
we obtain for the spectral density of the Laplacian 
\beq
\rho(m^2)= \frac{q}{2\pi} \frac{\sqrt{4(q-1) - (q-M)^2}}{q^2 -(q-M)^2} +O(1/n)
\eeq
with density support  $|(q-m^2)|< 2\sqrt{q-1}$.  Recently the $1/n$ correction
to the RRG spectral density has been evaluated \cite{Metz_2014}.

We observe two non-analyticities: the pole due to the zero modes in the
spectral density at $m^2=0$ and the branch points at $4(q-1) =(q-m^2)^2$.
The branch points seem to be interesting  and amount to criticality and nontrivial
"susceptibility"  $\frac{d^2\langle \log Z(m^2)\rangle}{d^2m^2}$ for $q=3$ at  points 
\(m^2_{crit}= 3 \pm 2\sqrt{2}\).   There is no such regime in the matrix model studied here: our graphs are planar whch drastically changes the critical behavior. We could ask the question whether this kind of distribution can occur in the double-scaled limit of the model, when we sum up over all genera  of graphs with the weight $N^{2-2\times{\rm genus}}$, close to criticality at every genus. This is however beyond the scope of our paper.

    \subsection{Analogy with QCD matrix model}
    
    Let us consider two-dimensional QCD, that is,
    fermions interacting with the quantum gauge fields. The partition function is the determinant
    of the Dirac operator averaged over  gauge fields $\langle \det (\hat{D}(A) +iM)\rangle_{YM,g_{YM}}$
    and it can be approximated by the large $N$ matrix model (see \cite{verbaarschot2000random} for a review) which has the
    following interpretation. The ground state is assumed to be populated by the instantons and antiinstantons
    hosting the  fermion  zero modes. The Wishart  matrix in the matrix model represents the Dirac operator in the basis of zero modes or speaking a bit
    differently the overlap of zero modes which get collectivized.
    
    In our case we have fermions interacting with gravity instead of the gauge field,
    determinant of the massive Laplace operator instead of the determinant of  Dirac operator and we do not 
    distinguish chirality. Therefore we have an analogue of the "instantons without antiinstantons" situation
    where each tree is the analogue of instanton hosting zero mode of the graph Laplacian.
    The number of the trees due to the index theorem coincides with the number of 
    zero modes hence like in QCD we could have a picture of the overlap of zero modes
    upon switching on gravity.
    Our finding for massive fermions interacting with gravity suggests that the formation 
    of the multiinstanton clusters in 2d  QCD can be
    expected as a function of the ratio of the gauge coupling constant and fermion mass.
    
    Fermionic condensates considered in this study have many similarities with the gluino condensates in SYM. There are still subtle problems concerning the evaluation of gluino condensates  due to the factorization of the higher topological correlators. It would be interesting to investigate the factorization issue 
    for the higher fermionic correlators in our case.

\section*{Acknowledgements}

We thank J. Bouttier, P. di Francesco, S. Komatsu, M. Mari\~{n}o, S. Nechaev,  D. Serban, A. Sportiello and N. Terziev for discussions. We are especially grateful to I. Kostov for illuminating comments. 
F.L.-M. is grateful for hospitality to organisers of the Varna ICMS-2022 workshop (organised by the International Center for Mathematical Sciences in Sofia
 and supported by the Simons Foundation). The research of V.K. was supported in part by the National Science Foundation under Grant No. NSF PHY-1748958. V.K. thanks the Interdisciplinary Scientific Center Poncelet (CNRS UMI 2615), where a part of the work was performed, for kind hospitality.
 A.G. is thankful to the Basis Foundation grant No. 20-1-1-23-1. The work of V.M. was funded by the "Basis" Foundation grant No. 22-1-1-42-3 and RFBR grants No. 21-52-52004 and No. 20-01-00644.

\appendix

\section{Combinatoric explanation of the Kirchoff-Tutte  matrix-tree theorem}
\label{sec:matrix-tree}
\paragraph{Theorem:}  The number of spanning trees of a connected loopless graph \(G \) is equal to the  determinant of \(\Delta'(G)\)  - the Laplacian  of the graph without it's last column and row.
The Laplacian itself is defined as:
\begin{equation}
\Delta=-\mathbb{Q}+A
\end{equation}
as in \eqref{logzeta}.

Let \(K_{k,\alpha}(G)\) be the (oriented) incidence matrix of a  graph  \(G\) with oriented edges labeled as \( \alpha=1,\dots,E_{G}\), and vertices labeled as \(\quad k=1,\dots n\), constructed as follows:
\begin{equation}
K_{k,\alpha}=\begin{cases}  1\,, & \text{ if the edge \(\alpha\) points from vertex \(k\) outside } \\
 -1\,, & \text{ if the edge \(\alpha\) points from vertex \(k\) inside } \\
0\,, & \text{otherwise.} 
\end{cases}
\end{equation}
It is easy to see that \(\Delta_{ij}=\sum\limits_{\alpha}K_{i,\alpha}K_{i,\alpha},  \) or \(\Delta=KK^{T} \).  Moreover this equatily "survives"\ on the level of the first minor - \(\Delta'=K'K'^{T} \), where \(K'\) is the Kirchhoff matrix with erased last line. 
\\

The proof is based on Cauchy-Binet formula: For two rectangular \(m\times M\) matrices \(F\)  and \(H\), we have
\begin{equation}
\det (FH)=\sum_{S_{m}\in \mathcal{S}_M} \det{F_{S_m}}\det{H_{S_m}}
\end{equation}
where by \(\det F_{S_n}\) we denote one of the maximal \(m\times m\) minors of the matrix \(F\) (and similarly for \(H)\). The sum goes over all \(\binom{M}{m}\) such minors.

Applying it to \(L'=K'K'^{T} \)  we write 
\begin{equation}
\det (\Delta')=\sum_{S_{m}\in \mathcal{S}_M} \det{K'_{S_m}}\det{K'^{T}_{S_m}=\sum_{S_{m}\in \mathcal{S}_M} (\det{K'_{S_m})^2}}
\end{equation}
Each matrix
\(K_{S_m}\)  is the Kirchoff  matrix of a subgraph obtained by cutting in \(G\)  all the rest of \(E_{G}-n\)  edges.

Next, we notice that the matrix \(K' \) is unimodular:  all its maximal minors are equal to \(0,\pm 1\).  It is zero if the subgraph has at least one cycle or is disconnected, and \(\pm 1\) otherwise (which means that it is a spanning tree). Indeed, each  maximal square submatrix  \(K_{S}\in K\)  is in fact the incidence matrix of the subgraph obtained by cutting all the edges corresponding to the erased columns of \(K\). Then \(K_S K_S^{T}\) is the laplacian on this subgraph. Then \( \det( K_SK_S^{T})\) is nonzero only if the graph is connected, or it has no loops (the connectivity and the looplessness are actually the same of such maximal subgraphs).  But in this case it is a spanning tree, and it is easy to see that for the spanning tree \(\det( K_S)=\pm 1\) (depending on the orientation): each column contains only one 1 and one (-1), and they can be always ordered in an upper-triangular way.

Then it is clear that 
\begin{equation}
\det (\Delta'(G))=\#\text{of spanning trees of }G.\\ 
\end{equation}       

{\it Quod erat demonstrandum!}

\section{Some properties of the spectral determinant}
\label{sec:spectral}

\begin{itemize}
    \item Define the spectral determinant of 
    undirected non-weighted graph G Laplacian L as
    \begin{equation}
       L(z,G)= \det(L-Iz)
    \end{equation}
    The matrix-forest theorem claims
    \begin{equation}
       L(z,G)=\sum_i(-1)^{i}a_i  z^{n-i}
    \end{equation}
    where $(n-i)$ is the number of trees in the forest and
    $a_i$ is the product of  number of nodes in all
    trees in $(n-i)$-component forest. 
     \begin{equation}
       \frac{d}{dz}L(z,G)|_{z=0}=a_{n-1}=n|t(G)|=\prod z_i
    \end{equation}
    where $z_i$ are roots of the Laplace polynomial. It 
    is the form of matrix-tree theorem.
    
    \item The reciprocity 
     \begin{equation}
       L(z,\bar{G^p})=(-1)^{n-1}L(np-z,G)
    \end{equation}
     \begin{equation}
       t(K^p -G)=(sp)^{s-n-2}L(sp,G)
    \end{equation}
    where $\bar{G^p}$ is graph p-complemented to G,
    $s$ is the number of nodes in the full graph $K^p$. 
    
    \item The complementarity
    \begin{equation}
    zL(z,G_1^pG_2)=(z-n_1p-n_2p)L(z-n_2p,G_1)L(z-n_2p,G_2)
    \end{equation}
    where $G_1,G_2$ are graphs with the number of nodes $n_1,n_2$
    
    \item The spectral determinant can be generalized a bit if we attribute the weights for the nodes $x(v_i)$ (contrary to the standard weights for links $\omega(v_i,v_j)$) forming the weight vector $\vec{x}$. The generalization of the matrix-forest and 
    matrix-tree theorems does exist in this case and reads
    as \cite{postnikov}
     \begin{equation}
      z^{-1}(-1)^{n-1} \det(M(x,G)-Iz)={\cal{F}}(z,x,G)
    \end{equation}
    where $M(x,G)=(m_{ij}=-x(v_j)\omega(v_i,v_j)) $ is the Laplacian
    matrix of the graph dressed by the node's degrees of freedom 
    and ${\cal{F}}(z,x,G)$ is 
    the so-called forest volume of G. 
    If $x(v_i)=1$ it 
    reduces to the matrix-forest theorem.
    \begin{equation}
       (-1)^{n-1}L(-z,G)={\cal{F}}(z,1,G)
    \end{equation}
    \end{itemize}

\section{Elliptic functions}
\label{app:ell}

Here we collect useful integrals and relations for elliptic functions we use.

\subsection{Some integrals}
\label{app:sint}
Let us give more details on taking the integral \eq{Gasint}. 
We assume that \(b<a<c\). For the first two terms it reduces to elementary functions,
\begin{equation}
\int_b^ady\frac{9 c M^2+y}{(x-y) \sqrt{(a-y) (y-b)}}=\frac{\pi  \left(-\sqrt{(a-x) (b-x)}+9 c M^2+x\right)}{\sqrt{(a-x) (b-x)}}
\end{equation}
 whereas the last one contains the complete elliptic integral of 3-rd kind:
\beq\int_b^a\frac{dy}{( x-y)}  \frac{y-2c}{\sqrt{(a-y)(y-b)(c- y)}}
=-\frac{2K\left(\frac{a-b}{c-b} \right)}{\sqrt{c-b}}+(x-2c)
\frac{2\Pi\left( \frac{a-b}{x-b},1-p \right)}{(x-b)\sqrt{c-b}}
\eeq
Combining them together we find the result \eq{Gres1} for $G(x)$.

\subsection{Relations used for integration of elliptic functions}

The relations we use to take the integral \eq{phii1} read:
\label{app:eint}
\begin{equation}
    \begin{split}
      n
   (n-1)^{k/2} \sqrt{\frac{l}{n}-1} \Pi (n|l)&=  -\frac{(k n+n+2) (n-1)^{k/2} \sqrt{\frac{l}{n}-1} (K(l) (l-n)+n E(l))}{(k+1) (k+3) n (l-n)} + \\
   &+ \dfrac{d}{dn} \left(\frac{2 (n-1)^{\frac{k}{2}+1} \left(\frac{2}{k+1}+n\right) \sqrt{\frac{l}{n}-1} \Pi (n|l)}{k+3} \right)
   \\
   (n-1)^{k/2}
   \sqrt{\frac{l}{n}-1} \Pi (n|l)&= -\frac{(n-1)^{k/2} \sqrt{\frac{l}{n}-1} (K(l) (l-n)+n E(l))}{(k+1) n (l-n)}+
   \\
   &+\dfrac{d}{dn}\left( \frac{2 (n-1)^{\frac{k}{2}+1} \sqrt{\frac{l}{n}-1} \Pi (n|l)}{k+1} \right)
    \end{split}
\end{equation}

\section{Expansions at small $M$ and $p$}
\label{sec:smallMp}

We have to order $M^2$ (dropping also terms of higher order than listed in $L$ and $p$)
\beqa && \nn
    a=M^2 (\frac{27 \left(25551 L^2+3 \left(38879+5778 \sqrt{2} \pi \right) L+3072 \pi ^2+44738 \sqrt{2} \pi +129502\right) p^2}{256 \pi
   ^2}
   \\  &&
   +\frac{27 \left(465 L^2+2 \left(841+255 \sqrt{2} \pi \right) L+4 \left(371+268 \sqrt{2} \pi +48 \pi ^2\right)\right) p}{16 \pi
   ^2}
   \\ \nn &&
   +\frac{27 \left(33 L^2+60 \left(1+\sqrt{2} \pi \right) L+8 \left(2+9 \sqrt{2} \pi +6 \pi ^2\right)\right)}{8 \pi ^2})
   \\ \nn &&
   +M
   \left(-\frac{3 \left(1455 \sqrt{2} L+768 \pi +2869 \sqrt{2}\right) p^2}{64 \pi }-\frac{3 \left(51 \sqrt{2} L+48 \pi +80 \sqrt{2}\right)
   p}{4 \pi } \right.
   \\ \nn && 
   \left.
   -\frac{3 \left(3 \sqrt{2} L+6 \pi +2 \sqrt{2}\right)}{\pi }\right)
   \\ \nn &&
   +2
\eeqa
and
\beqa
    &&
    b=M^2 (\frac{27 \left(-7317 L^2+\left(8052 \sqrt{2} \pi -96055\right) L+3072 \pi ^2+14392 \sqrt{2} \pi -215062\right) p^2}{256 \pi
   ^2}
   \\ \nn &&
   +\frac{27 \left(-87 L^2+54 \left(4 \sqrt{2} \pi -33\right) L-4 \left(943+\sqrt{2} \pi -48 \pi ^2\right)\right) p}{16 \pi ^2}
   \\ \nn &&
   +\frac{27
   \left(-3 L^2+4 \left(6 \sqrt{2} \pi -35\right) L-8 \left(22+11 \sqrt{2} \pi -6 \pi ^2\right)\right)}{8 \pi ^2})
   \\ \nn &&
   +M \left(\frac{3
   \left(198 \sqrt{2} L-384 \pi +1589 \sqrt{2}\right) p^2}{32 \pi }+\frac{3 \left(3 \sqrt{2} L-24 \pi +67 \sqrt{2}\right) p}{2 \pi }+\frac{42
   \sqrt{2}}{\pi }-18\right)
   \\ \nn &&
   -2
\eeqa
and
\beqa \nn
    &&
    c=M^2 (\frac{27 \left(35535 L^2+3 \left(59487+7730 \sqrt{2} \pi \right) L+3072 \pi ^2+67074 \sqrt{2} \pi +219742\right) p^2}{256 \pi
   ^2}
   \\  &&
   +\frac{27 \left(537 L^2+6 \left(347+97 \sqrt{2} \pi \right) L+4 \left(467+348 \sqrt{2} \pi +48 \pi ^2\right)\right) p}{16 \pi
   ^2}
   \\ \nn &&
   +\frac{27 \left(33 L^2+60 \left(1+\sqrt{2} \pi \right) L+8 \left(2+9 \sqrt{2} \pi +6 \pi ^2\right)\right)}{8 \pi ^2})
   \\ \nn &&
   +M
   \left(-\frac{9 \left(853 \sqrt{2} L+256 \pi +2439 \sqrt{2}\right) p^2}{64 \pi }-\frac{9 \left(21 \sqrt{2} L+16 \left(3 \sqrt{2}+\pi
   \right)\right) p}{4 \pi }
   \right.
   \\ \nn && \left.
   -\frac{3 \left(3 \sqrt{2} L+6 \pi +2 \sqrt{2}\right)}{\pi }\right)
   \\ \nn &&
   +4 p^2+4 p+2
\eeqa

\section{Lambert function and brane insertion}

\label{sec:lambert}

\subsection{Two examples with Lambert function}
In this Section we shall comment on the reason for the Lambert function
to appear in our double scaling limit. To preclude the arguments below recall our
setting. We start with fermions on the fixed graph and represent 
the massive determinant as the weighted sum over the separated trees in the forest.
Then we switch on the gravity  the interacting system itself selects the preferable state depending on the
point at the two-dimensional parameter space $(m,\lambda)$. Hence in any representation
of our model we need the boundary creation operator in some form yielding
the boundaries of the trees.

Before turning to our model let us describe two models where the Lambert
function has emerged in the very similar context. First, consider the setup
discussed in \cite{okuyama} where the perturbation of the 2d topological gravity 
via the boundary creation operator has been considered. The motivation of that study
was as follows. The boundary matrix model for JT gravity found in \cite{sss} involves the
important contribution from the replica wormholes providing the 
interaction of baby universes. It was suggested in \cite{okuyama} to substitute 
the standard replica trick with  useful integral representation for
quenched free energy $\langle\log Z\rangle$. It is equivalent to the replica representation 
with particular analytic continuation to $n\rightarrow 0$.

In this representation the quenched free energy reads 
\beq
\langle\log Z\rangle- \log \langle Z\rangle = - \int_0^{\infty} \frac{dx}{x} [ e^{-\tilde{Z}x} - e^{- \langle Z\rangle x} ]
\eeq
and involves the function $e^{\tilde{Z}(x)}$ which is the generating function for 
connected correlators $\langle Z^n\rangle_c$. These correlators  provide the multiple replica
boundary contributions in JT gravity partition function.
It has been also identified as the operator
creating the peculiar space-time D-brane \cite{Marolf_2020} introduced  in the context 
of replica wormholes. Remarkably if we consider the Airy
limit of the Gaussian model in genus zero this generating function can be evaluated exactly.
The exact boundary creating operator $\hat{Z}$ in topological gravity introduced 
in \cite{Moore:1991ir} is applied to the partition function of topological gravity
\beq
\langle Z^n\rangle_c= (\hat{Z})^n\cal{F}
\eeq
where
\beq
\hat{Z} =g_s \sqrt{\frac{\beta}{2\pi}}\sum_{k}\beta^k \partial_k
\eeq
and derivative is taken with respect to k-th time.
If one restricts oneself to the genus zero the Lambert function emerges 
as the generating function for boundary creating operator in topological gravity
at genus zero in the Airy limit. This is the
first role of the Lambert function.

The second role of Lambert function is important as well \cite{okuyama}. It is familiar
in the topological string context that the insertion of the brane shifts the closed
moduli in background which is usually written in symbolic relation $Z_{closed}(\vec{t'}) =Z_{brane}Z_{closed}(\vec{t})$
(see, for instance \cite{Aganagic_2005}). In the case under consideration this equation reads 
\beq
\tilde{Z}(x)= {\cal F}(t_k) - e^{-\hat{Z}x}{\cal F}(t_k) =
{\cal F}(t_k) - {\cal F}(t_k')
\eeq
hence we expect that "Lambert brane" amounts to the closed moduli shift. This argument turns
out to be true and the KdV flows in topological gravity yield the simple derivation of this shift \cite{okuyama}.
To this aim consider the case when only two first times in hierarchy are switched on $t_k=0, k\leq 2$.
The relevant solution to the KdV hierarchy in this case depends on two times $t_0,t_1$ and 
before the brane insertion it reads
\beq
u(t_0,t_1)= \frac{t_0}{1-t_1}
\eeq
where $u=\partial_0^2{\cal F}$. The shift of the moduli can be seen from the 
string equation which can be written for KdV in terms of Gelfand-Dikii polynomials $R_k$ as
follows
\beq
u(t)=\sum_k t_k R_k
\eeq
If there are only two non-vanishing times upon the shift  
string equation 
can be brought into the Lambert form \cite{okuyama}
\beq 
z=We^W
\eeq
for 
\beq
W=\frac{\beta t_0}{1-t_1}-\beta v \qquad  z= \frac{x}{1-t_1}e^{\frac{\beta t_0}{1-t_1}}
\eeq
Hence the second role of Lambert function - shift of the background 
via the brane insertion when only
two first times in integrable hierarchy are switched on.

The second model enjoying the Lambert function is the topological A model string on $CP_1$
coupled to topological gravity.
The theory has  two equivalent dual representations \cite{nekrasov} : a) all genus topological
A type string on $CP_1$ supplemented by the first gravitational
descendant of Kahler form and b) the massive two-dimensional 
fermions with specific gravity induced action. 
According to the topological string framework the inclusion of descendent 
corresponds to the insertion of some brane similar to the topological gravity above. 
The  first representation of Lambert function as the brane creation operator is not explicitly
known in this model hence we focus on the second role -- shifting the moduli. This aspect has been
identified in \cite{nekrasov} in explicit form.

It goes as follows. The model is solved in terms of the spectral curve which is the
 sphere with two marked points. The cut on the $C$ plane lies between
 $x_{\pm}= v\pm \Lambda$.
 The filling fraction for fermions is defined as $a=\hbar R$ and
 plays the role of closed moduli. This theory similarly to the topological gravity case 
 involves only two times $t_1,t_2$
and it was found in \cite{nekrasov} that instead of KdV hierarchy for pure gravity here 
the semiclassical limit of Toda hierarchy does the job.
 When $t_2=0$ we have no gravity
 descendant of Kahler form  and the relation $v=a$ holds. When $t_2\neq 0$
 solution requires the proper matching condition at the ramification
 points which yields shift of the closed moduli.  
 The Lambert function $W(z)$ enters  
via the following matching conditions for moduli shift
\begin{equation}\label{shift}
    v-a= W(t_2e^{t_1+at_2})
    \end{equation}
    \begin{equation}
    \log\Lambda = t_1 +2t_2 a- 1/2 W(-16t_2^2 e^{2(t_1+2t_2a)})
    \end{equation}
As in the previous case we interpret this relation as an effect of insertion of the "Lambert brane"
and its backreaction on gravity.

Integrability  implies the fermionic  representation of the partition function
\cite{nekrasov}
 
\begin{equation} \label{ferm}
    Z(t_1,t_2,R)=\langle R|e^{-\frac{J_1}{\hbar}} e^{\hbar^{-1} (t_2 W_3 + t_1 W_2)}
   e^{-\frac{J_{-1}}{\hbar}} |R\rangle
\end{equation}
where $|R\rangle$ is the state with  $U(1)$ charge $R$, $\omega$ is the coordinate on the cylinder,
$t_1=\log \Lambda$.
The harmonics of the $U(1)$ current are defined as 
\beq{}
J_k=\sum_{r} :\tilde{\psi}_r \psi_{r+k}:
\eeq
The generators of the  $W_{1+\infty}$ algebra are expressed in terms of fermions as follows
\beq
W_{k+1}= - \frac{\hbar^k}{k+1} \oint d\omega (\tilde{\psi}[(D+1/2)^{k+1} - (D-1/2)^{k+1}]\psi)
\eeq
where $D= \omega \partial _{\omega}$. 
Hence we observe that the fermions are effectively massive and have a non-standard kinetic term 
involving the second derivative.

The time $t_2$ is coupled to the unusual second derivative term  induced by coupling to gravity and coinciding with the zero mode of the $W_3$
generator of the  $W_{1+\infty}$ algebra. Upon bosonization it can be expressed
as the $(\partial \phi)^3$ term in the action for the chiral boson and  has an interpretation
as a cut-and-join operator. Such terms are familiar in many models (see for 
instance \cite{Dijkgraaf_2002} for the relevant discussion) and the coefficient in front of this term has an
interpretation of the string coupling.

This theory has one more interpretation which actually was the initial
one for the authors of \cite{nekrasov}.
The model can be viewed as the $D=4$ abelian $U(1)$
${\cal N}=2$ SYM theory in the $\Omega$ background when the leading 
gravitational correction to the prepotential is taken into account. That is, in the UV
the prepotential involves only two times
$${\cal {F}}_{UV}= t_1 \Tr\;\Phi^2 + t_2 \Tr\;\Phi^3$$
where the second term is induced by the coupling to gravity.
Naively there are no instantons in the
abelian theory but they do emerge when the coupling to gravity
is switched on and the theory is asymptotically free with a non-perturbatively
generated IR scale. Presumably this theory can be considered as the
worldvolume theory on the inserted brane.

\subsection{Lambert in the forest}

From two examples above we see that the Lambert function plays the role of the generating
function for multiple boundary insertions and simultaneously shifts the closed moduli. In our case 
we indeed need  tree boundary creation operators, however according to the matrix-forest
theorem the boundary of each tree enters with the volume of the tree. Hence the second $CP^1$ example is more similar to our case since the effect of the
gravity descendant of the Kahler class involves the volume factor indeed. 
As in that case we will observe the shift of the closed moduli.

Let us first note that  the Lambert function $W(x)$ in \eqref{j1(z)1} in our notation
is just the $z$ variable or more precisely its rescaled version $t$ (see \eq{redefs}), 
\begin{equation}
    t=-W(-\mu^{-1}e^{c_0\frac{g-g_c}{\mu}})
\end{equation}
where $W(x)$ obeys the equation $W(x)e^{W(x)}=x$ 
and numerical factor $c_0= \frac{1}{256 \pi}$.
Recall that near the
critical point the area of the surface $A$ behaves as $A\propto \frac{1}{g-g_c}$
hence we have $e^{-\frac{1}{MA}}$ in the argument of  the Lambert function.
This can be compared with the standard  instanton exponent 
$e^{-\frac{1}{g_s A}}$ which implies the suppression of the instanton effects 
at large $N$ since $g_s=1/N$. In our case it seems that $g_s\propto M$
which implies the lack of large $N$ suppression. Notice that the factor like
$e^{-\frac{1}{g_s A}}$ was discussed for  the instanton contributions in 2d YM theory.

Let us turn now to two roles of the Lambert function observed in the previous examples.
First, one can ask whether we have an effective brane generating multiple boundaries on the
worldsheet like in the Airy limit of topological gravity at genus zero.
It is a well known fact that the insertion of $\det(M-x)$ in the matrix model for the
type B topological string corresponds to the insertion of the FZZT brane 
at point $x$ of the spectral curve of the matrix  model. 
In our case the massive  determinant $\det(\Delta +M)$ involved in our partition function
according to the matrix-forest theorem seems to play such a role. Indeed its 
expansion in mass  provides multiple trees on the worldsheet.
Hence we expect that we have the effective brane insertion as well
and the mass $M$ provides the position of the  insertion. Certainly this
point deserves further clarification.

Secondly, similarly to  the example of the topological A model string on $CP^1$ we  can look for the
equation describing the shift of the closed moduli via 
the insertion of the brane. The shift has to be proportional to the 
deformation parameter $M$. The equation entering our one-gap solution
\begin{equation} \label{edge}
  p= 16Mz = MW(x)  
\end{equation}
plays this role.
When $M=0$ and $p\rightarrow 0$ the
elliptic curve  degenerates into the marked sphere while
the insertion of the "Lambert brane" at point $x$ yields the 
modified condition (\ref{edge}) which is the analog of (\ref{shift}).

Notice  that there are several other precise examples relating
non-critical strings and topological strings. In particular the $c=1$ string 
is described as the topological string on a conifold \cite{Ghoshal_1995} and minimal models
coupled to 2d gravity were argued to be described via a topological string on 
a particular Calabi-Yau manifold. The account
of the gravity descendants in the minimal model corresponds to the adding of the
B-branes into the B model geometry and insertion points are the open moduli.

The Lambert function has a finite radius of convergence   $x_0=\frac{1}{e}$ in the series
representation $$W(x)=\sum_{n=1}n^{n-1}\frac{x^n}{n!}\ .$$ The inspection of the radius
of convergence in our case yields the condition for the area $A$ in the critical regime
\beq{}
A>\frac{1}{M\log M}
\eeq
which implies the validity of the approximation in this regime only. The meaning 
of a possible transition at this radius deserves special study.

\bibliographystyle{unsrt}
\bibliography{ERexp}

\end{document}